\documentclass[onecolumn]{aastex631}

\def    \apjl  		{\rm {ApJL}}
\def    \apj  		{\rm {ApJ}}
\def    \mnras  	{\rm {MNRAS}}
\def    \araa  		{\rm {ARA\& A}}
\def    \apjl  		{\rm {ApJL}}

\def	\cm		{\,{\rm {cm}}}
\def	\K		{\,{\rm K}}
\def	\g		{\,{\rm {g}}}
\def	\mum	{\,{\mu \rm{m}}}

\def \bea {\begin{eqnarray}}
\def \ena {\end{eqnarray}}

\def	\B	{{\rm B}}

\def	\C	{{\rm C}}

\def	\cm	{\,{\rm cm}}
\def	\d	{{\rm d}}

\def	\D	{{\rm D}}

\def	\erg	{\,{\rm erg}}

\def	\g	{\,{\rm g}}

\def	\G	{{\rm G}}

\def	\H	{{\rm H}}

\def	\N	{{\rm N}}

\def	\s	{\,{\rm s}}

\def	\V	{{\rm V}}

\accepted{2021}

\begin{document}
\shorttitle{Polarized dust emission in 30 Doradus}
\shortauthors{Tram et al.}

\title{SOFIA observations of 30 Doradus: I - Far-Infrared dust polarization and implications for grain alignment and disruption by radiative torques}

\author{Le Ngoc Tram}
\altaffiliation{Current address: Max-Planck-Institut f\"ur Radioastronomie, Auf dem H\"ugel 69, 53-121, Bonn, Germany}
\affiliation{Stratospheric Observatory for Infrared Astronomy, Universities Space Research Association, NASA Ames Research Center, MS 232-11, Moffett Field, 94035 CA, USA}
\correspondingauthor{Le Ngoc Tram}
\email{nle@mpifr-bonn.mpg.de}

\author{Thiem Hoang} 
\affiliation{Korea Astronomy and Space Science Institute, Daejeon 34055, South Korea}
\affiliation{Korea University of Science and Technology, 217 Gajeong-ro, Yuseong-gu, Daejeon, 34113, South Korea}

\author{Enrique Lopez-Rodriguez} 
\affiliation{Kavli Institute for Particle Astrophysics and Cosmology (KIPAC), Stanford University, Stanford, CA 94305, USA}

\author{Simon Coud\'e}
\affiliation{Stratospheric Observatory for Infrared Astronomy, Universities Space Research Association, NASA Ames Research Center, MS 232-11, Moffett Field, 94035 CA, USA}

\author{Archana Soam}
\affiliation{Stratospheric Observatory for Infrared Astronomy, Universities Space Research Association, NASA Ames Research Center, MS 232-11, Moffett Field, 94035 CA, USA}

\author{B-G Andersson}
\affiliation{Stratospheric Observatory for Infrared Astronomy, Universities Space Research Association, NASA Ames Research Center, MS 232-11, Moffett Field, 94035 CA, USA}

\author{Min-Young Lee}
\affiliation{Korea Astronomy and Space Science Institute, Daejeon 34055, South Korea}

\author{Lars Bonne}
\affiliation{Stratospheric Observatory for Infrared Astronomy, Universities Space Research Association, NASA Ames Research Center, MS 232-11, Moffett Field, 94035 CA, USA}

\author{William D. Vacca}
\affiliation{Stratospheric Observatory for Infrared Astronomy, Universities Space Research Association, NASA Ames Research Center, MS 232-11, Moffett Field, 94035 CA, USA}

\author{Hyeseung Lee}
\affiliation{Korea Astronomy and Space Science Institute, Daejeon 34055, South Korea}

\begin{abstract}
Located in the Large Magellanic cloud and mostly irradiated by a massive-star cluster R$\,$136, 30 Doradus is an ideal target to test the leading theory of the grain alignment and rotational disruption by RAdiative Torques (RATs). Here, we use publicly available polarized thermal dust emission observations of 30 Doradus at 89, 154, and 214$\,\mu$m using SOFIA/HAWC+. We analyse the variation of the dust polarization degree ($p$) with the total emission intensity ($I$), the dust temperature ($T_{\d}$), and the gas column density ($N_{\rm H}$) constructed from {\it Herschel} data. The 30 Doradus complex is divided into two main regions relative to R$\,$136, namely North and South. In the North, we find that the polarization degree first decreases and then increases before decreasing again when the dust temperature increases toward the irradiating cluster R$\,$136. The first depolarization likely arises from the decrease of grain alignment efficiency toward the dense medium due to the attenuation of the interstellar radiation field and the increase of the gas density. The second trend (the increase of $p$ with $T_\d$) is consistent with the RAT alignment theory. The final trend (the decrease of $p$ with $T_\d$) is consistent with the RAT alignment theory only when the grain rotational disruption by RATs is taken into account. In the South, we find that the polarization degree is nearly independent of the dust temperature, while the grain alignment efficiency is higher around the peak of the gas column density and decreases toward the radiation source. The latter feature is also consistent with the prediction of the rotational disruption by RATs.

\end{abstract}

\section{Introduction\label{sec:intro}}
Grain alignment induces the polarization of background starlight as well as the polarization of thermal dust emission. Dust polarization induced by aligned dust grains is widely used to map magnetic fields (see e.g., \citealt{2007JQSRT.106..225L}). A leading theory describing grain alignment is based on RAdiative Torques (hereafter RATs), which arises from the interaction of an anisotropic radiation field with irregular dust grains \citealt{1976Ap&SS..43..257D}; \citealt{Draine1996}; \citealt{2007MNRAS.378..910L}; see \citealt{2015psps.book...81L} and \citealt{2015ARA&A..53..501A} for recent reviews). According to the RAT alignment theory, the alignment efficiency of dust grains depends on the radiation field and the local gas properties (e.g., \citealt{2021ApJ...908..218H}; \citealt{2021ApJ...907...93S}). Gas collisions tend to damp the grain rotation and randomize the grain orientation, whereas RATs act to spin-up and align dust grains. Therefore, toward the center of a dense cloud, the degree of grain alignment by RATs decreases due to the attenuation of the interstellar radiation field (ISRF) and the increase of gas density. This results in the decrease of the dust polarization degree ($p$) toward the center of molecular clouds (MCs), i.e., higher gas column density $N_{\H}$ (see e.g., \citealt{2021ApJ...908..218H}). Numerous observations toward MCs report the decrease of the polarization degree with $N_{\H}$, which favors the RAT alignment theory (\citealt{2008ApJ...674..304W}; \citealt{2014A&A...569L...1A}; \citealt{2020ApJ...905..157V}). Note that the tangling of magnetic fields is also suggested to produce the decrease of the polarization with increasing $N_{\H}$ (e.g., \citealt{1992ApJ...389..602J}). 

With the advance of high resolution polarimetric facilities (JCMT/POL2, ALMA, SOFIA/HAWC+), one can now observe the polarization induced by dust grains in proximity of an embedded source (e.g., a protostar) where the effect of stellar radiation becomes dominant over that of the ISRF. As a result, toward the embedded source, the RAT alignment theory predicts the increase of the dust polarization with increasing the radiation emission intensity (or dust temperature $T_{\d}$, \citealt{2021ApJ...908..218H}). Numerical modeling (\citealt{2020ApJ...896...44L}) and numerical simulations (\citealt{2016A&A...593A..87R}) of dust polarization using the RAT alignment theory report a monotonic increase of the polarization degree of thermal dust emission with increasing the radiation intensity (or $T_{\d}$). Such a correlation of $p$ with $T_{\d}$ is an important feature predicted by the RAT alignment theory. As a result, observing dust polarization around a strong radiation source is crucial to test the RAT alignment theory (see e.g., \citealt{2019ApJ...873...87M}; \citealt{2021AJ....161..149S}).

Interestingly, polarimetric observations reveal that the degree of thermal dust polarization does not always increase with the radiation intensity (or dust temperature, $T_{\rm d}$) and decrease with increasing $N_{\H}$, as implied by the RAT alignment theory. For example, \textit{Planck} observations at 850 $\mu$m toward four molecular clouds, including Aquila Rift, Cham-Musca, Orion, and Ophiuchus in the Gould Belt cloud, showed that the polarization degree decreases for $T_{\rm d}>19~\K$ (see \citealt{2020A&A...641A..12P}). Shorter wavelength polarimetric measurements at 500 $\mu$m from the air-balloon BLASTPol toward the Vela C also show a similar decrease at $T_{\d}\geq 19\K$ (see \citealt{2016ApJ...824..134F}). Further shorter wavelength observations by the High-resolution Airborne Wide band Camera Plus (HAWC+) instrument (\citealt{2018JAI.....740008H}) aboard the Stratospheric Observatory for Infrared Astronomy (SOFIA) toward the molecular cloud Ophiuchus A at 89 and 154 $\mu$m also reported the decrease of the polarization degree for $T_{\rm d}\geq 25-32\K$ and $N_{\H} \leq 10^{22}\cm^{-2}$ (see \citealt{2019ApJ...882..113S} and \citealt{2021ApJ...906..115T}). For the case of $\rho$ Ophiuchus A, the peak of the dust temperature is close to the central source (a B-type star), while the gas column density peaks further away. Thus, the decrease of the dust polarization degree toward high dust temperatures and low gas column density (toward the source) could not be explained by the loss of grain alignment due to weak radiation intensity or enhanced collisional randomization as in the case of starless cores (\citealt{2008ApJ...674..304W}; \citealt{2014A&A...569L...1A}).

\cite{2019NatAs...3..766H} realized that large grains cannot survive once being exposed to a strong radiation field owing to the Radiative Torque Disruption (RATD) mechanism (see \citealt{2020Galax...8...52H} for a review). The RATD mechanism causes the fragmentation of large grains into many smaller species when the centrifugal stress induced by suprathermal rotation due to RATs exceeds the tensile strength of the grain material, which is determined by the binding energy that hold their constituents together. RATD is more efficient for large grains because the RAT efficiency increases with the grain size (\citealt{2007MNRAS.378..910L}; \citealt{2008MNRAS.388..117H}). The depletion of large grains due to RATD is expected to result in the decrease of the polarization degree of the thermal dust emission because large grains dominate the polarized emission at long wavelengths.

The first numerical modeling of dust polarization that considers both grain alignment and rotational disruption by RATs was performed in \cite{2020ApJ...896...44L}. They found that the dust polarization degree increases monotonically with $T_{\d}$ when grains are aligned by RATs and the RATD effect is not taken into account. In the presence of RATD, the polarization degree first increases and then decreases when the dust temperature becomes sufficiently large (i.e., the corresponding radiation strong enough) to activate the RATD effect. The critical temperature for the RATD effect depends on the local gas density and grain tensile strength that depends on the grain structure. Their modeling results could successfully reproduce the anti-correlation $p-T_{\d}$ trend at 850$\,\mu$m observed by \textit{Planck} (\citealt{2020A&A...641A..12P}). As a result, the variation of $p$ with $T_{\d}$ observed toward an intense radiation source becomes an important test for the RAT alignment and RATD theory. The first detailed analysis of the $p-T_{\d}$ relation is carried out in \cite{2021ApJ...906..115T} for the SOFIA/HAWC+ observations from Ophiuchus A cloud at 89 and 154 $\mu$m. The authors found the anti-correlation of $p-T_{\d}$ for sufficiently large $T_{\d}$ and show that it could be reproduced by the RAT alignment and RATD. \cite{2021ApJ...908...10N} also found the anti-correlation of $p-T_{\d}$ toward the proximity of the LkH$\alpha$ 101 star in the Auriga-California cloud. These studies reveal the importance of RATD that needs to be taken into account together with RAT alignment to interpret dust polarization data observed toward an intense radiation source. Moreover, observational data also allow us to constrain the physical properties and characteristics of dust grains, such as shape, internal structure, mineralogy, helicity and size distribution.

\begin{figure*}
    \centering
    \includegraphics[width=0.7\textwidth]{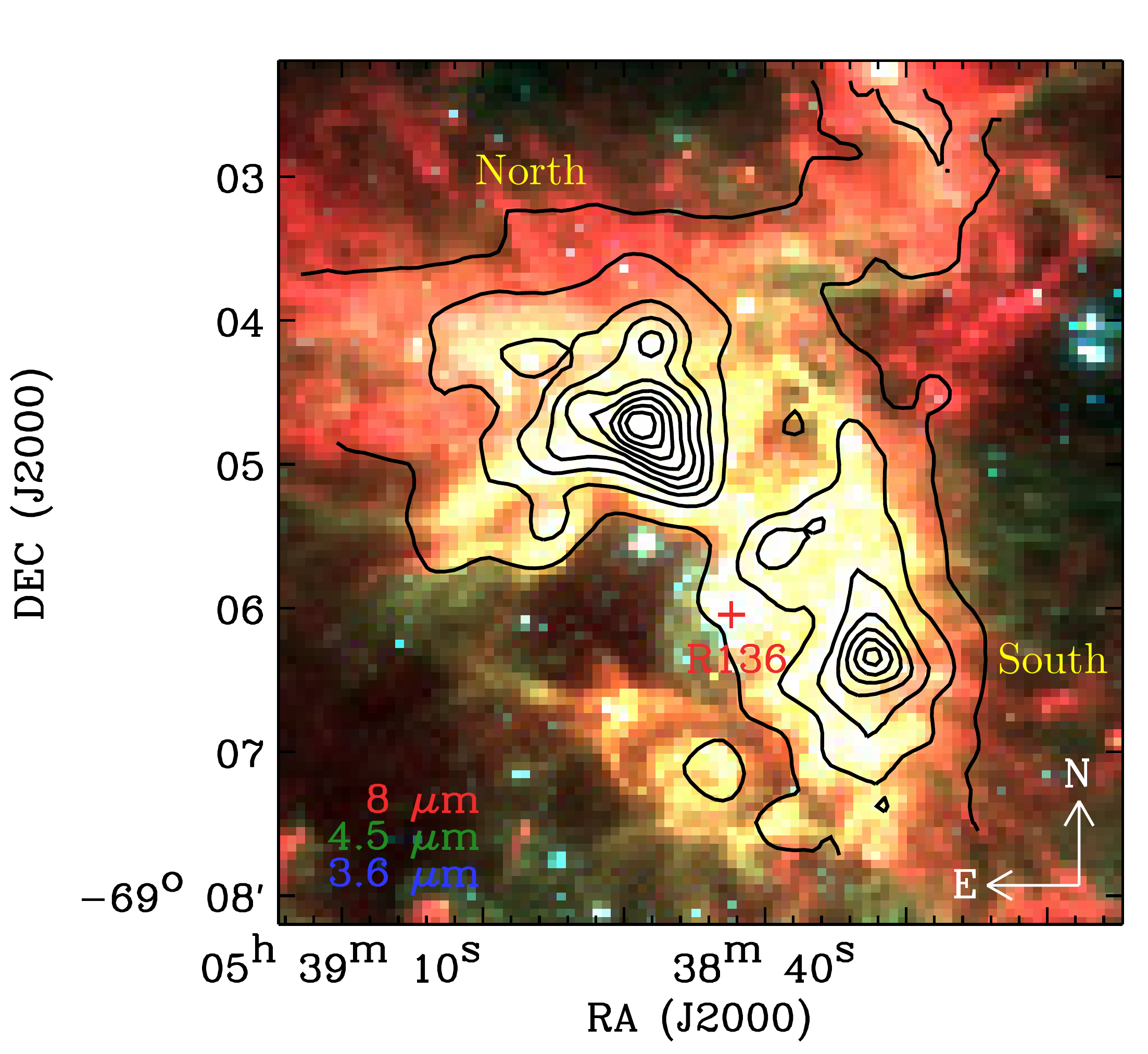}
    \caption{Background is a RGB composite image with R: 8$\,\mu$m, G: 4.5$\,\mu$m, and B: 3.6$\,\mu$m. The black contours show an example of the FIR observations at 154$\,\mu$m with SOFIA/HAWC+. There are two main regions in 30 Dor: North and South in relative to the massive star cluster, R$\,$136. The data analyzed in this work locate in the strong radiation field.}
    \label{fig:RGB}
    \includegraphics[width=1.0\textwidth]{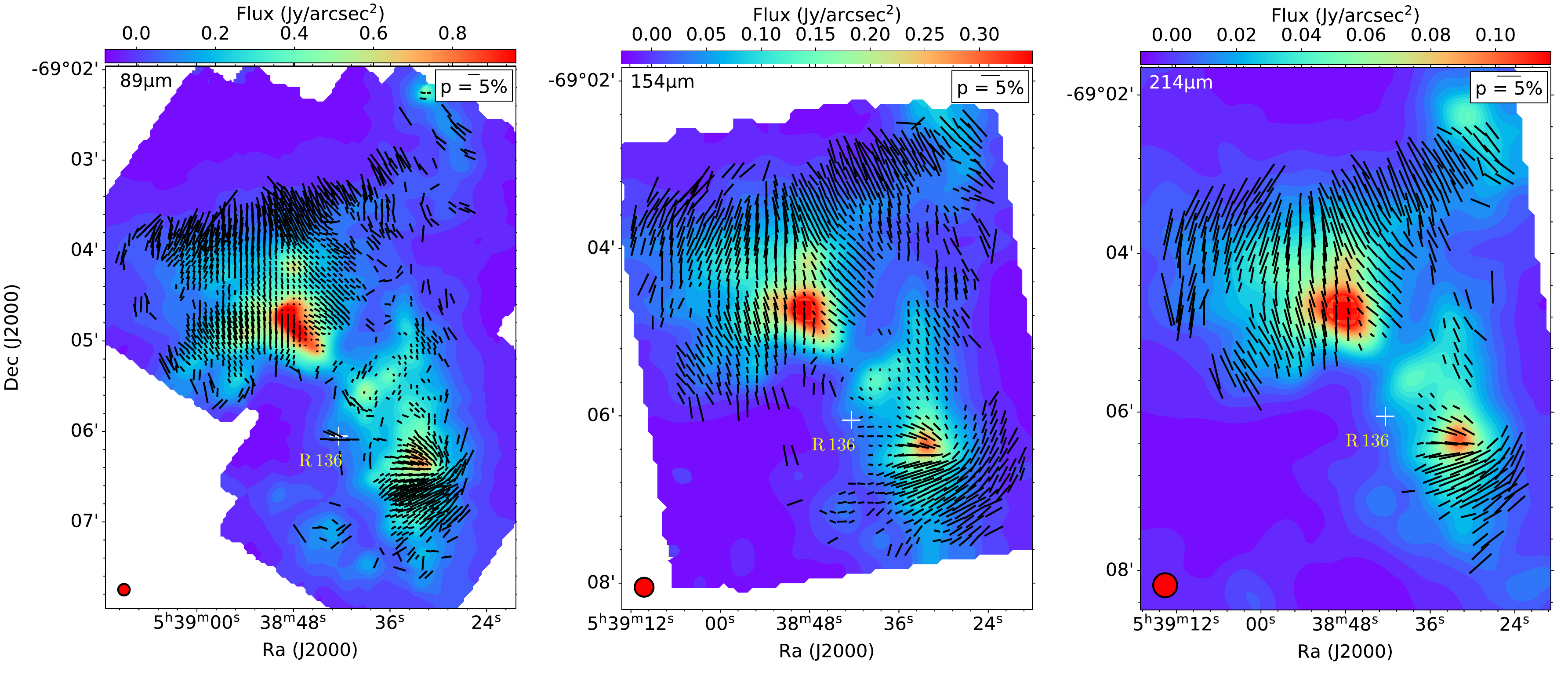}
    \caption{Polarization maps (E-vectors) of 30 Dor observed using SOFIA/HAWC+ at 89 (left), 154 (middle), and 214 $\mu$m (right). The background color is the total intensity (Stokes I). The vectors are selected within the thresholds of $I/\sigma_{I}\geq 100$ and $p/\sigma_{p}\geq 3$ (see Section \ref{subsec:obs_SOFIA}). The beam size (red circle) at the respective wavelengths and a 5\% polarization measurement (top-right) are shown. The location of R136 (white cross) is shown in every panel.}
    \label{fig:hawc_map}
\end{figure*}

In this paper, we will use the thermal dust polarization observed by SOFIA/HAWC+ at 89, 154, and 214 $\mu$m toward a massive star-forming cloud, 30 Doradus (hereafter 30 Dor), to test the RAT alignment and RATD mechanisms. 30 Dor is located in the Large Magellanic Cloud (LMC) with the distance of $\sim$ 50 kpc from us (e.g., \citealt{2011ApJ...739...27D}) and powered by a massive star cluster R$\,$136, with mass of $\simeq 5\times 10^{4} M_{\odot}$ (see \citealt{2009ApJ...694...84I} and references therein) and the bolometric luminosity of $7.8\times 10^{7}L_{\odot}$ (see \citealt{2011ApJ...731...91L} and references therein). Note that LMC is a low-metallicity galaxy ($Z\simeq 0.5Z_{\odot}$, \citealt{1982ApJ...252..461D}; \citealt{2008ApJ...679..310G}; \citealt{2016A&A...590A..36C}), with a low dust-to-gas ratio (e.g., $\sim$2-5$\times 10^{-3}$, see \citealt{2014ApJ...797...86R}; \citealt{2016A&A...590A..36C}), which makes the dust-shielding relatively low such that dust grains can be affected by the radiation field over a large scale (see Figure \ref{fig:RGB}). Thus, the 30 Dor cloud offers a valuable environment to test the physics of grain alignment and disruption by RATs. Toward this end, we will concentrate on the correlation of the polarization degree with the emission intensity, the dust temperature, and the gas column density. 

This paper is structured as follows. We first present the FIR polarimetric observations of 30 Dor observed by SOFIA/HAWC+ in Section \ref{sec:obs}. We then present (1) the distribution of the polarization angle and the polarization degree, (2) the relation of the polarization degree with the total intensity, the gas column density, and with dust temperature in Section \ref{sec:results}. The implications of the observational data for grain alignment and rotational disruption by RATs are presented in Section \ref{sec:discussion}. The summary of our main findings is presented in Section \ref{sec:summary}. 

\section{Observations of 30 Doradus} \label{sec:obs}
\subsection{SOFIA/HAWC+ multi-wavelength polarization}\label{subsec:obs_SOFIA}
Polarization of thermal dust emission from 30 Dor was observed by SOFIA/HAWC+ at three bands: C at 89$\,\mu$m, D at 154$\,\mu$m, and E at 214$\,\mu$m. The beams sizes (Full-Width-at-Half-Maximum, FWHM) are 7.8$''$, 13.6$''$, and 18.2$''$ at 89$\,\mu$m, 154$\,\mu$m, and 214$\,\mu$m, respectively. Data are publicly available under the Strategic Director's Discretionary Time (S-DDT) program (PI: Yorke, H., ID: 76$\_$0001), which observations were taken during the SOFIA New Zealand deployment in July 2018. Data have been presented by \citealt{2018arXiv181103100G} and no further data reduction has been performed.

For the linear polarization, the polarized intensity is defined as (\citealt{2018arXiv181103100G})
\bea
    I_{p} = \sqrt{Q^{2}+U^{2}}
\ena
with $Q$ the Stokes parameter Q, $U$ the Stokes parameter U. The association error is then
\bea
    \sigma_{I_{p}} = \left[\frac{(Q\sigma_{Q})^{2}+(U\sigma_{U})^{2}}{Q^{2}+U^{2}} \right]^{1/2}
\ena
with $\sigma_{Q}$ the error of Stokes Q, and $\sigma_{U}$ the error of Stokes U. 

The debiased polarization degree is calculated as
\bea
    p = \frac{100}{I}\sqrt{Q^{2}+U^{2}-\sigma^{2}_{Ip}}=100\frac{I_{p}}{I} ~~~(\%)
\ena
with $I$ the Stokes I parameter. The associated error is then
\bea \label{eq:sigmap}
    \sigma_{p} = p\left[\left(\frac{\sigma_{I_{p}}}{I_{p}}\right)^{2} + \left(\frac{\sigma_{I}}{I}\right)^{2}\right]^{1/2}
\ena
with $\sigma_{I}$ the error of Stokes I. 

Finally, the polarization angle and its error are given by
\bea
    \theta = \frac{1}{2}\arctan{\left(\frac{U}{Q}\right)} 
\ena
and
\bea
    \sigma_{\theta} = \frac{1}{2}\frac{\sqrt{(Q\sigma_{U})^{2}+(U\sigma_{Q})^{2}}}{Q^{2}+U^{2}}.
\ena

\begin{figure*}
    \centering
    \includegraphics[width=0.8\textwidth]{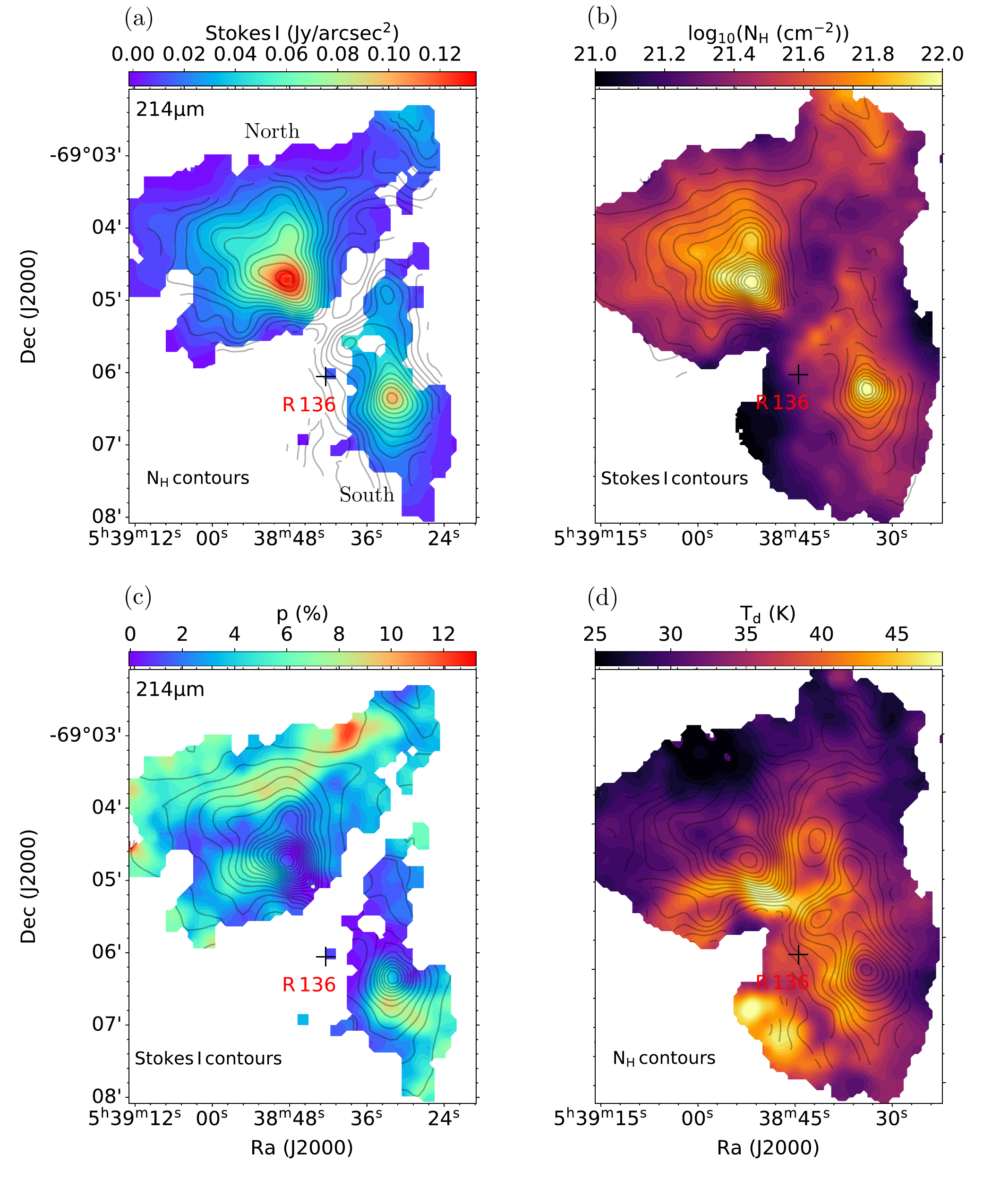}
    \caption{Panels(a,c): Maps of the total intensity and the polarization degree at 214$\,\mu$m after masking from the common sky positions in which all 3 bands in Figure \ref{fig:hawc_map} detected the signal. Panels(b,d): Maps of the gas column density and the dust temperature. The black cross locates the massive star cluster R$\,$136. The total intensity correlates well to the gas column density. The gas column density peak is offset from the dust temperature, whose peak is close to R$\,$136.}
    \label{fig:I_TdNH}
\end{figure*}

We then performed quality cuts to obtain statistically significant polarization measurements. The quality cuts on the polarization map are very conservative. In this work, we mask our HAWC+ maps by using two common thresholds. The first quality cut depends on the signal-to-noise ratio of the Stokes-I ($I/\sigma_{I}$), we adopt $I/\sigma_{I}\geq 100$, which allows $\sigma_{p}\simeq \sqrt{2}\times(I/\sigma_{I})^{-1} = 1.4\%$\footnote{This approximation is derived from Equation \ref{eq:sigmap} by assuming that $\sigma_{Q}=\sigma_{U}$}. The second threshold is the signal-to-noise ratio of the fractional polarization $p/\sigma_{p}\geq 3$. Figure \ref{fig:hawc_map} show the final map of the polarization vectors (E-vectors, not B-vectors). The length of the polarization measurements is referenced to $5\%$ of polarization degree. Data for band C (89$\,\mu$m), D (154$\,\mu$m) and E (214$\,\mu$m) is show from left to right, and the background color is the original total intensity (Stokes I). Figure \ref{fig:I_TdNH}(a,c) show the final maps of the stokes I and the polarization degree at 214$\,\mu$m as an example. The correlations to the gas column density are visualized by the black contours.

\subsection{Gas column density and dust temperature}
We constructed the maps of dust temperature, $T_{\d}$, and the gas column density, $N_{\H} \simeq N(\H+\H_{2})$, using \textit{Herschel} data at 100, 160, 250, 350 and 500 $\mu$m from HERITAGE (\citealt{2013AJ....146...62M}). The observations were registered and sampled to the 160 $\mu$m with a pixelscale of 3$''$ and FWHM = 11.4$''$, then a modified black-body function was used to fit to each of the line-of-sigh (LOS) SED with an assume spectral index $\beta$ = 1.62. This value was taken as the median from the range of $\beta \sim 1.5-1.8$ by \cite{2014ApJ...797...85G}, and $\beta=2.0$ used by \cite{2008AJ....136..919B}. 

\begin{figure*}
    \centering
    \includegraphics[width=0.32\textwidth]{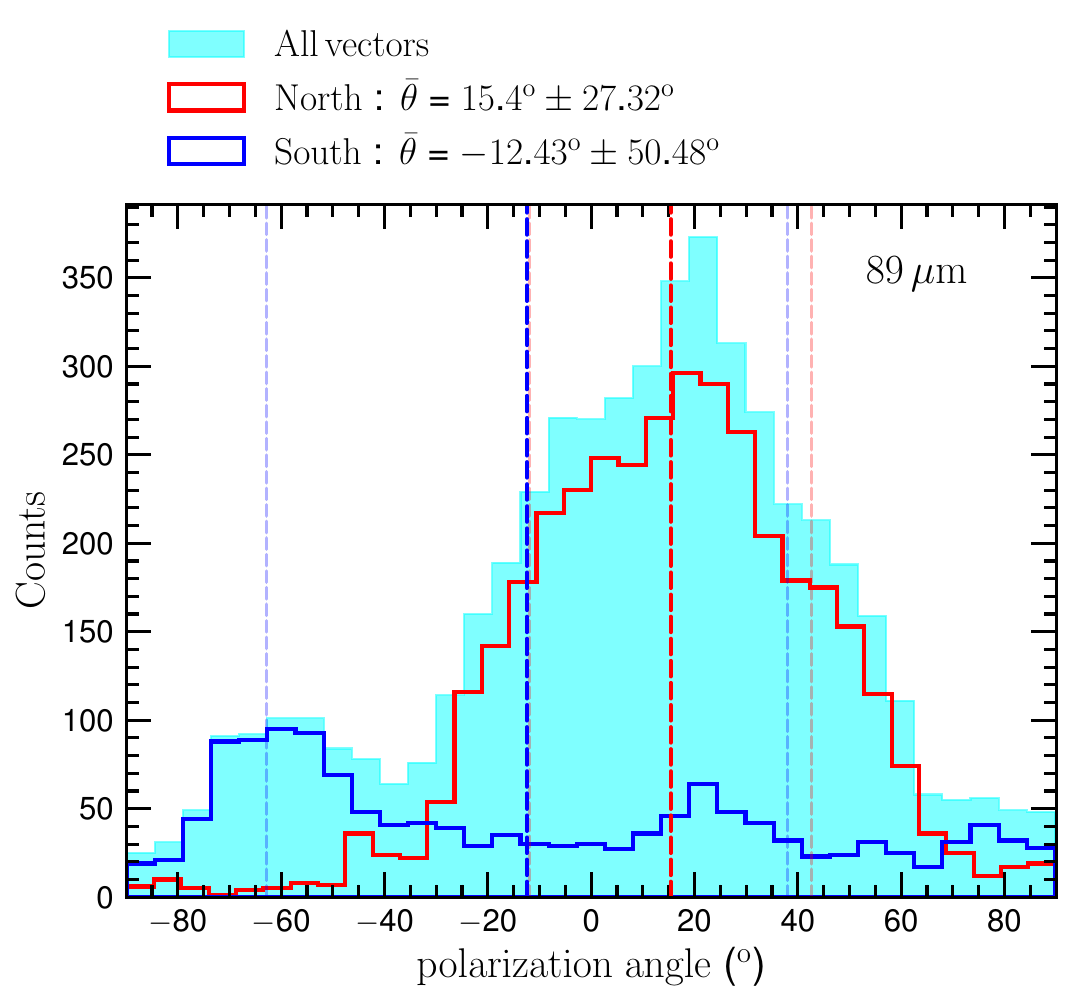}
    \includegraphics[width=0.32\textwidth]{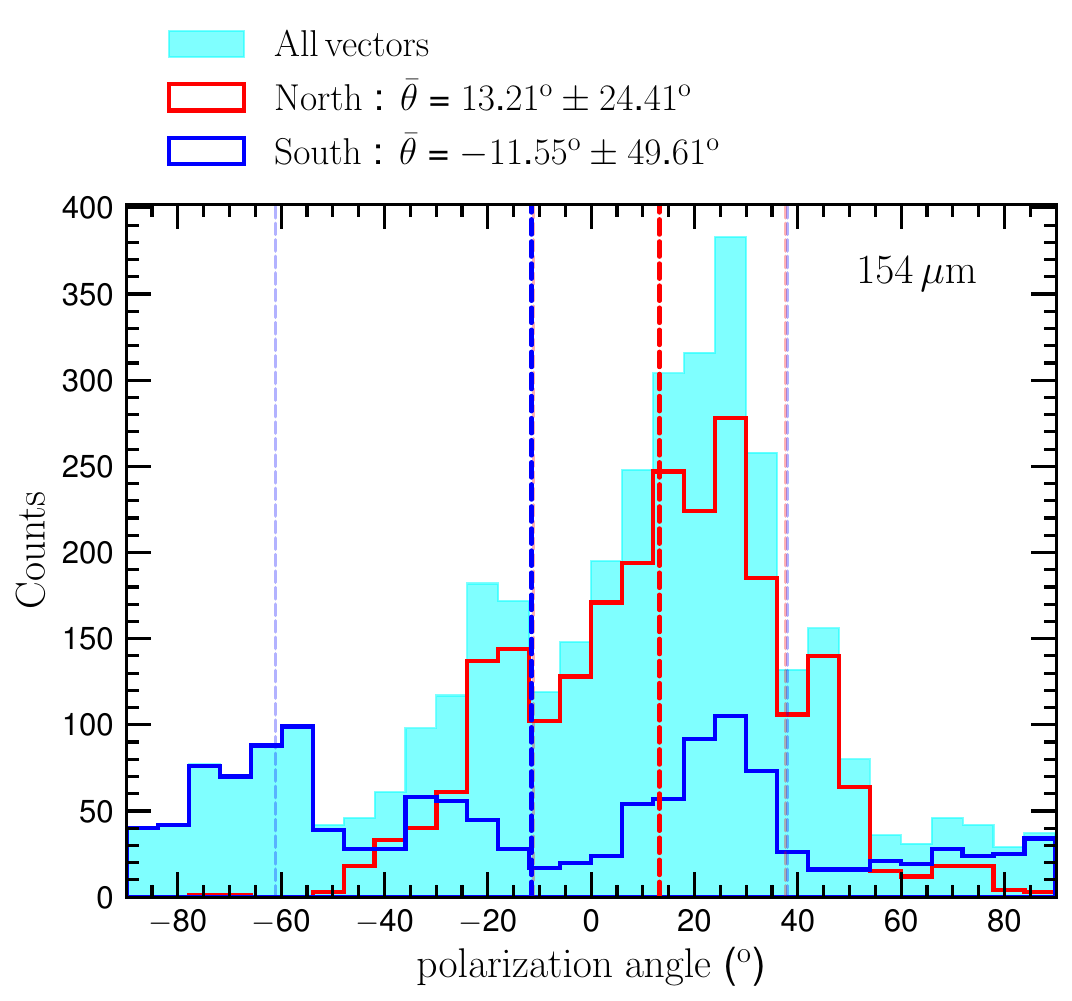}
    \includegraphics[width=0.32\textwidth]{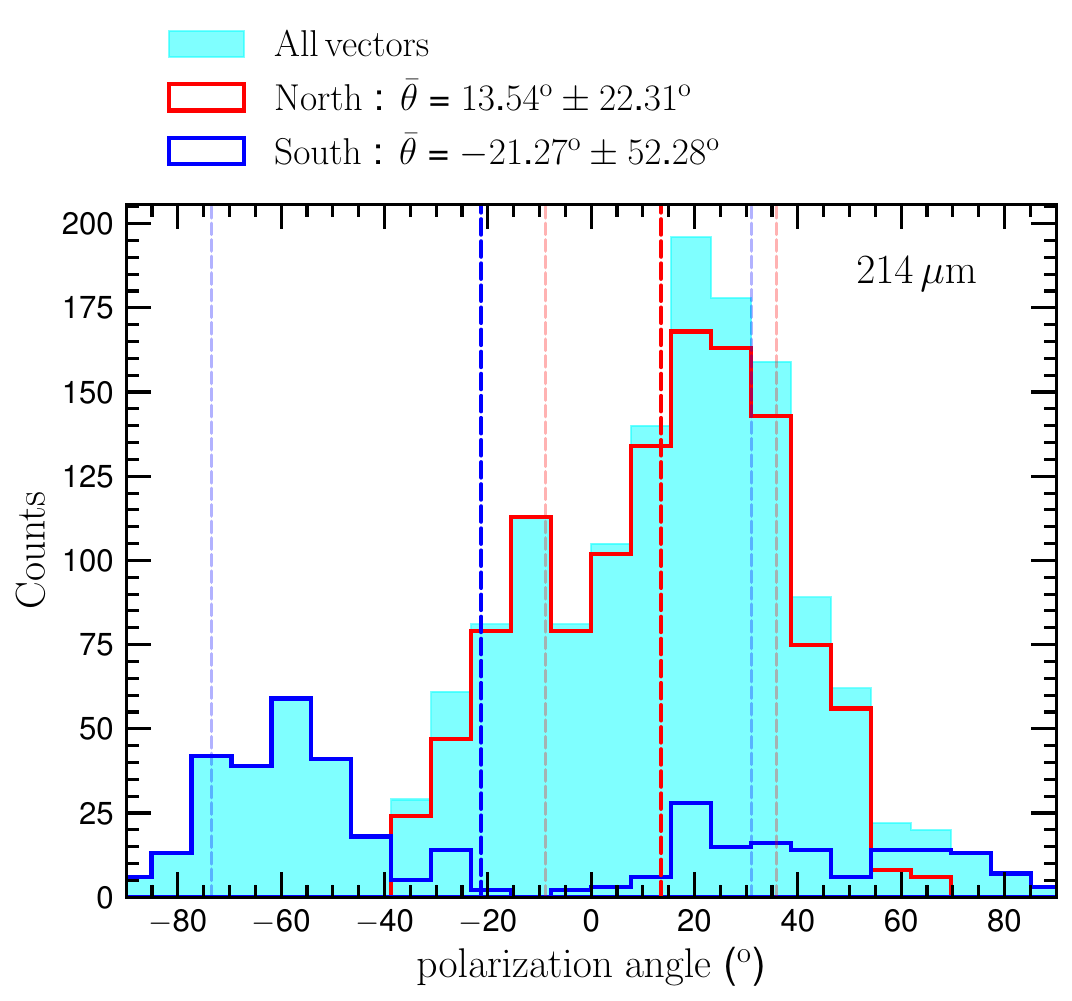}
    \includegraphics[width=0.32\textwidth]{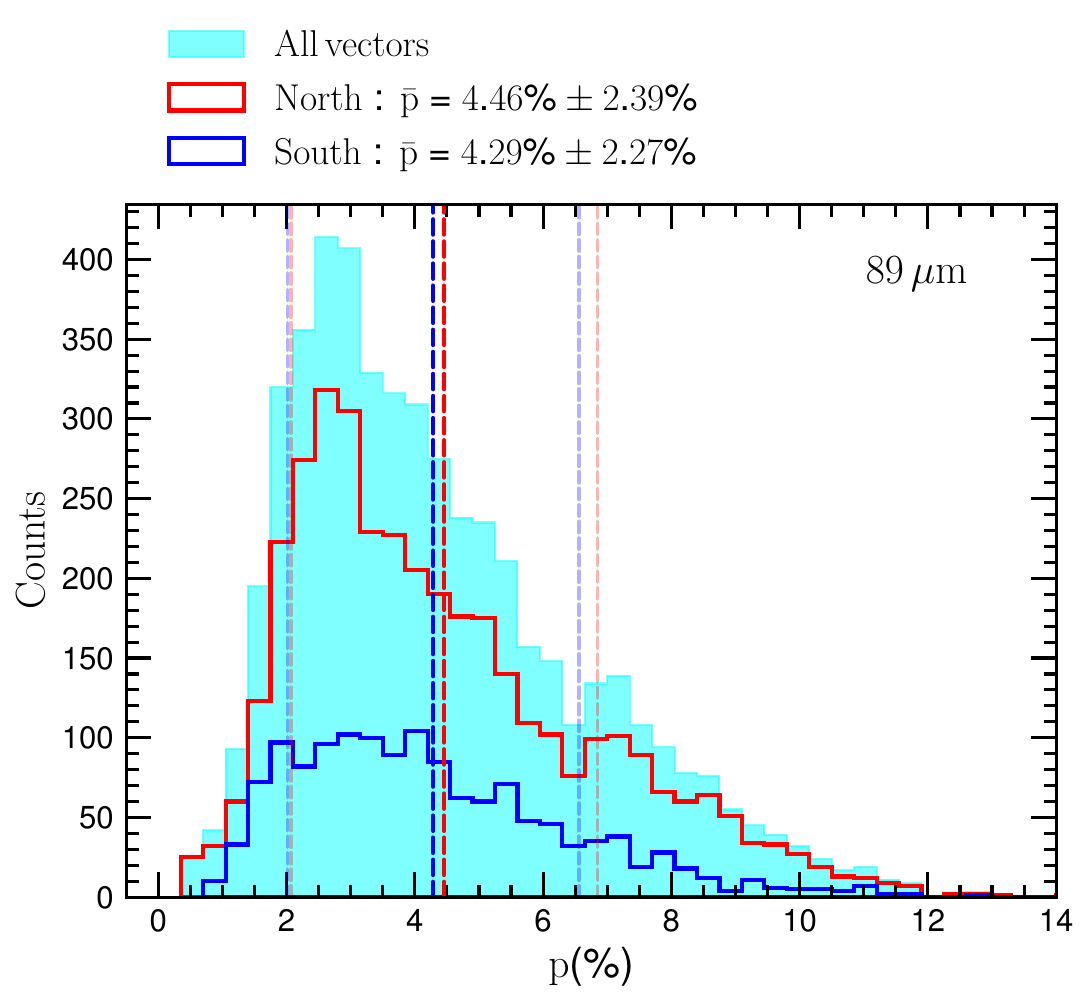}
    \includegraphics[width=0.32\textwidth]{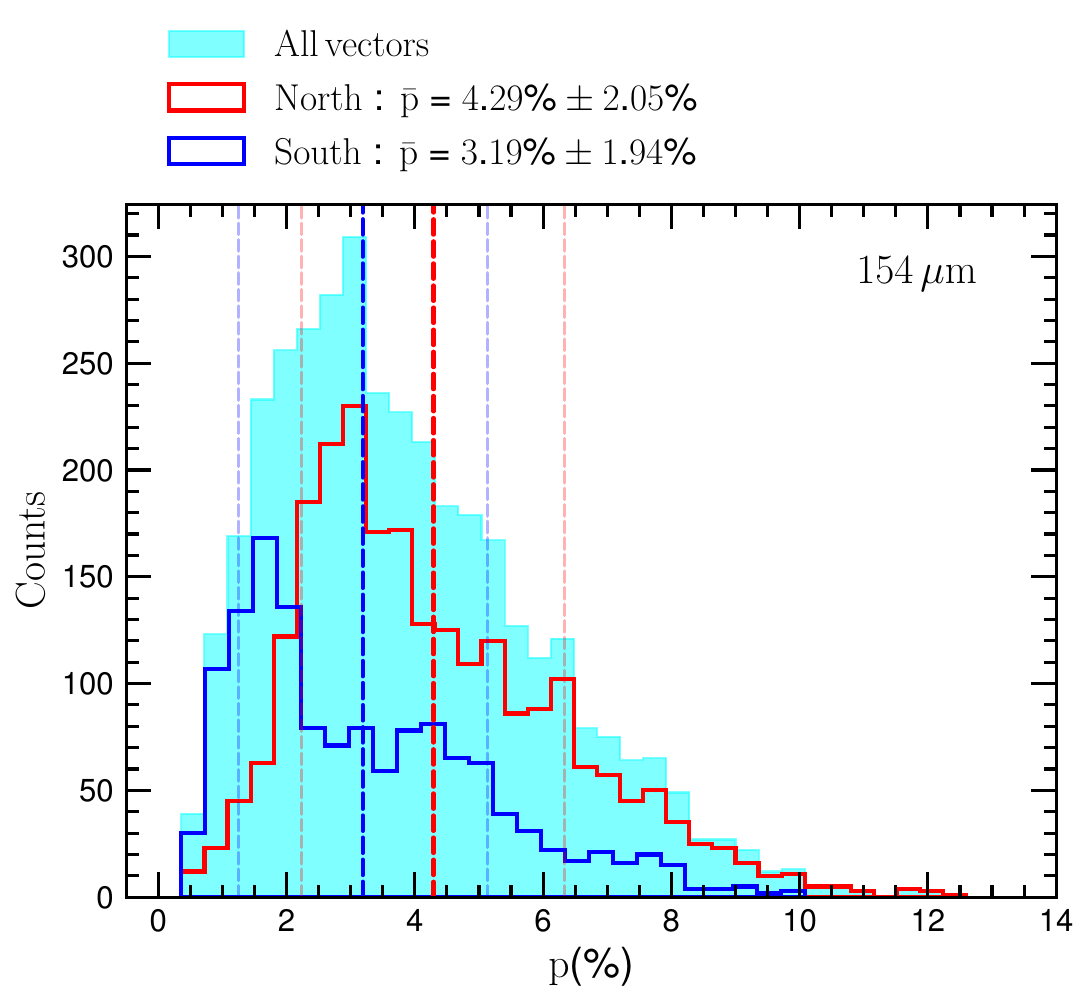}
    \includegraphics[width=0.32\textwidth]{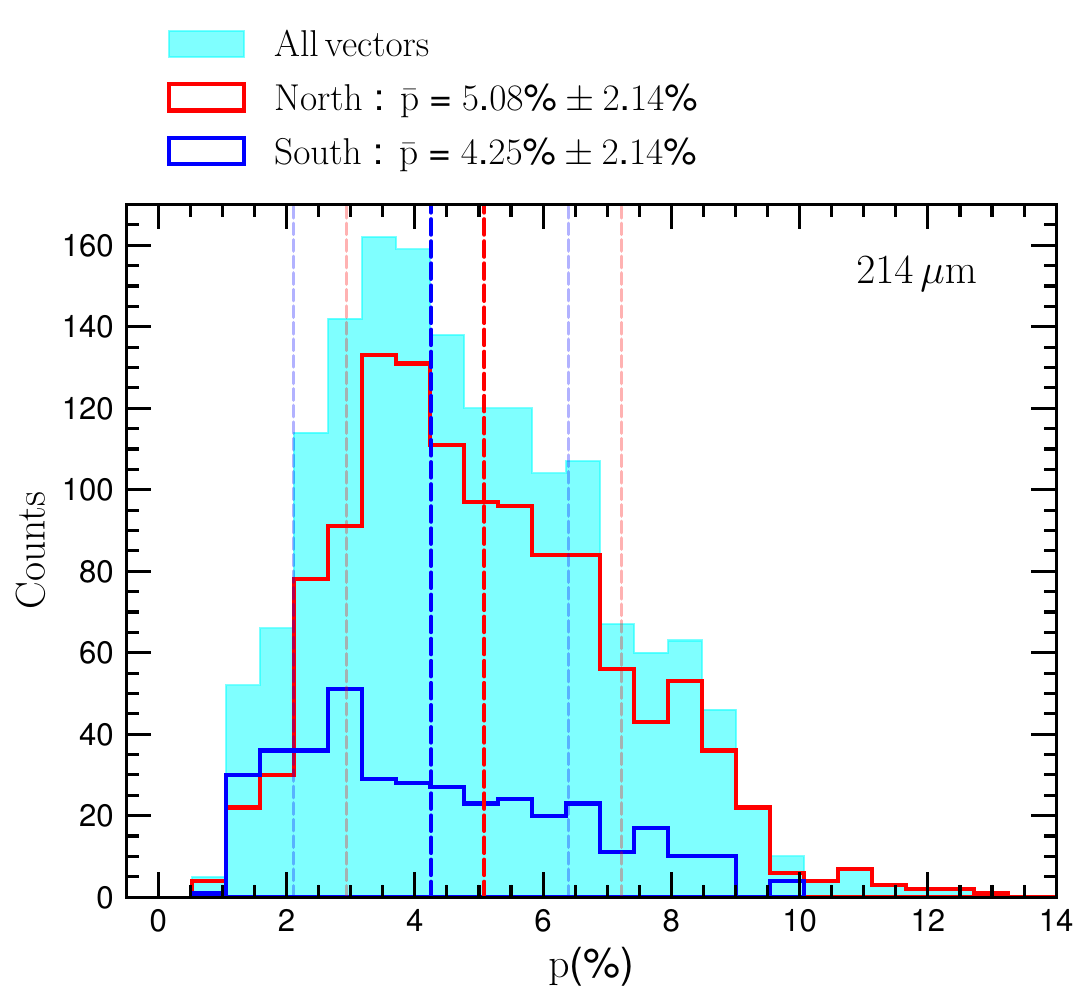}
    \caption{Distribution of the polarization angle (upper panels) and the polarization degree (lower panels) in 3 bands (from left to right). The cyan, red and blue are the distributions of the polarization measurement in the entire region, in the North and in the South regions (see Figure \ref{fig:hawc_map}), respectively. The vertical dashed red and blue lines are the weighted mean of the North and the South distributions. The corresponding thinner lines represent for the $1\sigma$ uncertainties.}
    \label{fig:pA_hist}
\end{figure*}

Panels (b,d) in Figure \ref{fig:I_TdNH} show the maps of the gas column density and dust temperature in 30 Dor. Toward R$\,$136, the dust temperature increases, while the gas column density first increases and then decreases after its peak, which is further away from the center. The correlations are visualized by the black contours in these maps. In addition, the dust temperature (gas column density) in the North is hotter (denser) than in the South.

\section{Results}\label{sec:results}
We now show the results analyzing the polarization degree of thermal dust emission in 30 Dor in terms of the total emission intensity, dust temperature and gas column density.

\begin{table}[]
    \centering
    \begin{tabular}{l|c c| c c} 
    \hline
    \hline
    {Bands} & \multicolumn{2}{c|}{$\theta^o(\pm 1\sigma)$} & \multicolumn{2}{c}{$p(\%)(\pm 1\sigma)$} \\
    {} & North & South & North & South \\
    \hline
        89$\,\mu \rm m$  & 15.4(27.32)  & -12.43(50.48) & 4.46(2.39) & 4.29(2.27) \\
        154$\,\mu \rm m$ & 13.21(24.41) & -11.55(49.61) & 4.29(2.05) & 3.19(1.94) \\
        214$\,\mu \rm m$ & 13.53(22.31) & -21.27(52.28) & 5.08(2.14) & 4.25(2.14) \\
    \hline
    \end{tabular}
    \caption{Mean value and uncertainties of the polarization angle ($\theta$) and the polarization degree ($p$) in the North and South regions within 3 bands.}
    \label{tab:fitting}
\end{table}

\subsection{Variation of polarization in 30 Dor}
The upper panels in Figure \ref{fig:pA_hist} show the distribution of the polarization angle of the thermal dust polarization in 30 Dor at 3 bands. We chose to use a position angle
span of -90 to +90 to center the position angles around the peak of the distribution. 
The variation of the polarization angle is very similar for the three bands. The distributions of all vectors are shown by the cyan area, which shows that the polarization vectors vary quite significantly, spanning from $-90^{\circ}$ to $90^{\circ}$, and peaks at around $20^{\circ}$ and $-60^{\circ}$. The first peak is characterized of the polarization in the North (see the red distribution). The second peak originates predominately in the South (see the blue distribution), which has a more random statistic. These distributions have the same bin width, which is chosen following the Freedman-Diaconis rule in python-package\footnote{https://docs.astropy.org/en/latest/visualization/histogram.html} from the all vectors. The weighted-mean values and the uncertainties have been calculated as shown by the dashed vertical thick and corresponding thin lines, whose values are listed in Table \ref{tab:fitting}.

The lower panels in Figure \ref{fig:pA_hist} show the distribution of the polarization degree in 3 bands. The distribution peaks at between $2\%$ and $4\%$. The red and blue histograms show the distributions in the North and South, respectively. The choice of the bin width is the same as above. The weighted mean and the associated $1\sigma$ uncertainties are showed by the dashed vertical thick and thin lines, whose values are also listed in Table \ref{tab:fitting}. Notably, the $\sigma$ is larger than the maximum associated uncertainty of $1.4\%$, thus the $1\sigma$ is intrinsically associated to changes in the source.

\subsection{Polarization degree versus total intensity}\label{sec:p_vsI}
\begin{figure*}
    \centering
    \includegraphics[width=1.0\textwidth]{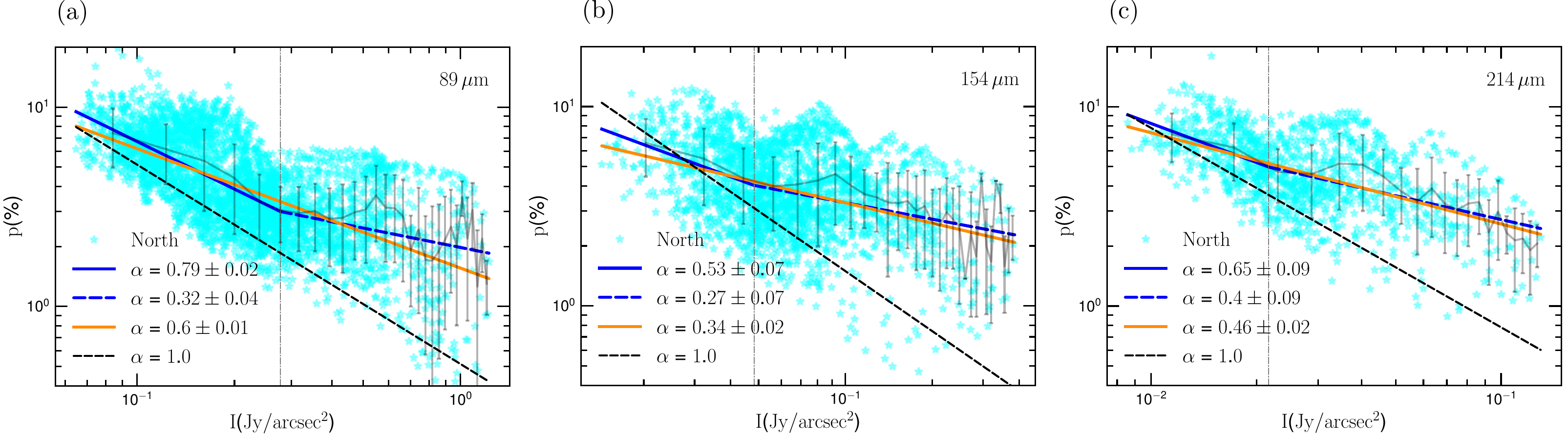}
    \caption{Relation of the polarization degree and the total intensity in 3 bands in the North region. The gray error bars are the mean-weighted within the bins of each data. A single power-law fitting is shown by the orange line, while the blue line is for a double power-law. In the latter case, the transitions are shown by the dashed vertical lines. As a comparison, the power-law with $\alpha=1$ is shown by the dashed black line.}
    \label{fig:pvsI_north}
    \includegraphics[width=1.0\textwidth]{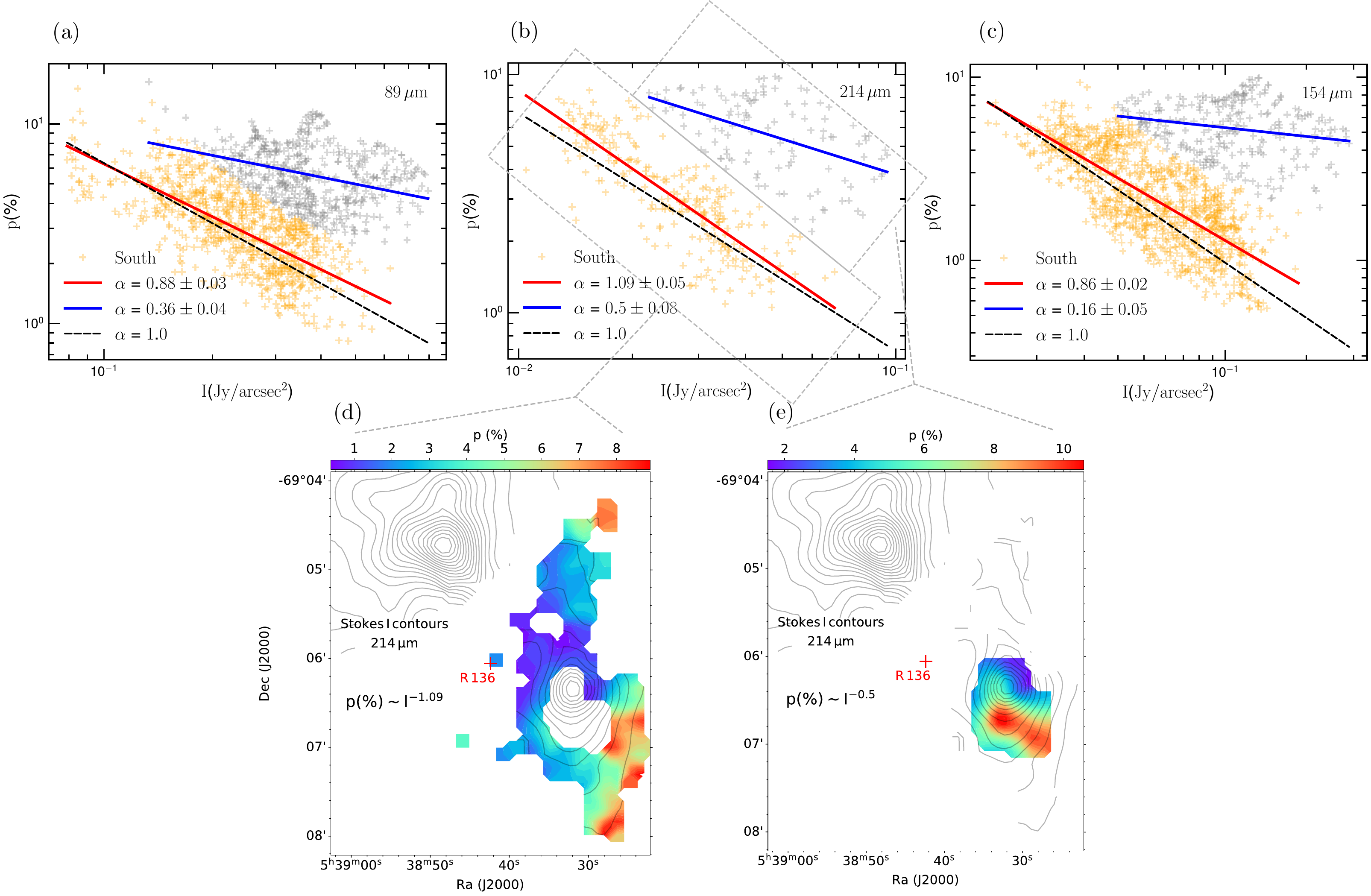}
    \caption{Similar to Figure \ref{fig:pvsI_north} but for the South. The data points likely group separately, hence we classify them into two different clusters by using \textsc{BayesianGaussianMixture} library in python-package. These clusters are colored by the gray and orange points in panels (a,b,c), which are followed by the power-law fittings. The slope differs from one to another cluster, which indicates a distinguished grain alignment efficiency between the two regions. Panels(d,e) visualize the spatial distribution of these clusters. The slope is shallower at around the peaked intensity and gas column density, whereas the polarization decreases steeply to higher intensity.}
    \label{fig:pvsI_south}
\end{figure*}
Figure \ref{fig:pvsI_north} shows the relation of the polarization degree to the total intensity for all 3 bands in the North region. We first fit the observational data with a single power-law function (solid orange line) using the \textsc{lmfit} python-package (\citealt{newville_matthew_2014_11813}), and compare it with the limiting case of $\alpha =1$ (black dashed line). The best fit indicates the slope of $\alpha=0.6\pm 0.01$ at 89$\,\mu$m is steeper than 0.5 (in the case the medium is fully turbulent, \citealt{1992ApJ...389..602J}) and the estimated slopes at 154$\,\mu$m ($\alpha=0.34\pm 0.02$) and 214$\,\mu$m ($\alpha=0.46\pm 0.02$) are shallower than 0.5. We then fit the data with a double power-law function (solid and dashed blue line) using the piecewise linear least square fit \textsc{pwlf} (\citealt{pwlf}). The best fits illustrate two distinct slopes, whose transitions are marked by the dashed dotted vertical lines. The transitions occur at 0.28$\pm 0.02\,\rm Jy/arcsec^{2}$, 0.05$\pm 0.01\,\rm Jy/arcsec^{2}$, and 0.02$\pm 0.003\,\rm Jy/arcsec^{2}$ at 89$\,\mu$m, 154$\,\mu$m, and 214$\,\mu$m, respectively. In the latter case, as the intensity increases, the polarization degree first declines rapidly with a steep slope of $\alpha>0.5$ (solid blue line), and then varies with a shallow slope with $\alpha<0.5$ (dashed blue line). The first steep slope is still shallower than $\alpha=1$ predicted by the model in which grains are only aligned in the outer envelope of the cloud (see e.g., \citealt{2021ApJ...908..218H}). We note that the slope difference gets smaller for longer wavelengths.

Figure \ref{fig:pvsI_south} shows the relation of the polarization degree to the total intensity for all 3 bands in the South region (panels a,b,c). The relations are complex, but the data appear to group into two different clusters. We then used the \textsc{BayesianGaussianMixture} library in \textsc{scikit-sklearn} python-package (\citealt{scikit-learn}) to classify the data points into 2 groups. Since the separation at 89$\,\mu$m is less obvious than at 154 and 214$\,\mu$m, the classification works well at these longer wavelengths and fails at 89$\,\mu$m. Thus, to make it consistent, we classify this case into 3 groups then merge two of them together. Each cluster is plotted in a different color and is fitted by a power law, as shown in the legend. The polarization degree is higher, and the slope is steeper from one cluster to another. 

To better understand the observed features, we show in panels (e) and (d) of Figure \ref{fig:pvsI_south} the spatial positions of these two clusters at 214$\,\mu$m as an example. The central region (colored area in panel e) seems to be shielded partially from the irradiation of R$\,$136, whereas the outer region (colored area in panel d) contains a near-side region which is directly irradiated by the source R$\,$136. Surprisingly (also differing from the North), the high polarization degree and shallower slope of the $p-I$ diagram is seen for the central region, located around the intensity peak (or the gas column density peak, see Figure \ref{fig:I_TdNH}a). Moreover, the polarization degree in the near-side region (close to R$\,$136) is much lower than that in the far-side region (lower and upper right corners; see color bars). The low polarization degree at higher radiation intensity is unexpected from the basic RAT alignment theory.

We note that the best-fits to the $p-I$ data only reflect the general trend and cannot describe accurately the underlying physics across the different regions. Indeed, as shown in Figure \ref{fig:pvsI_north}, there are many data points that appear to follow the steep slope of $\alpha=1$ toward high intensity. This can also be seen from Figure \ref{fig:I_TdNH}c that the polarization degree is minimum around the central radiation source, while it becomes higher at the farthest side from this source where the dust temperature is relatively low (see Figure \ref{fig:I_TdNH}b,c,d). Therefore, in the next section, we will perform additional analysis of the polarization with the gas column density and dust temperature.

\begin{figure}
    \centering
    \includegraphics[width=0.45\textwidth]{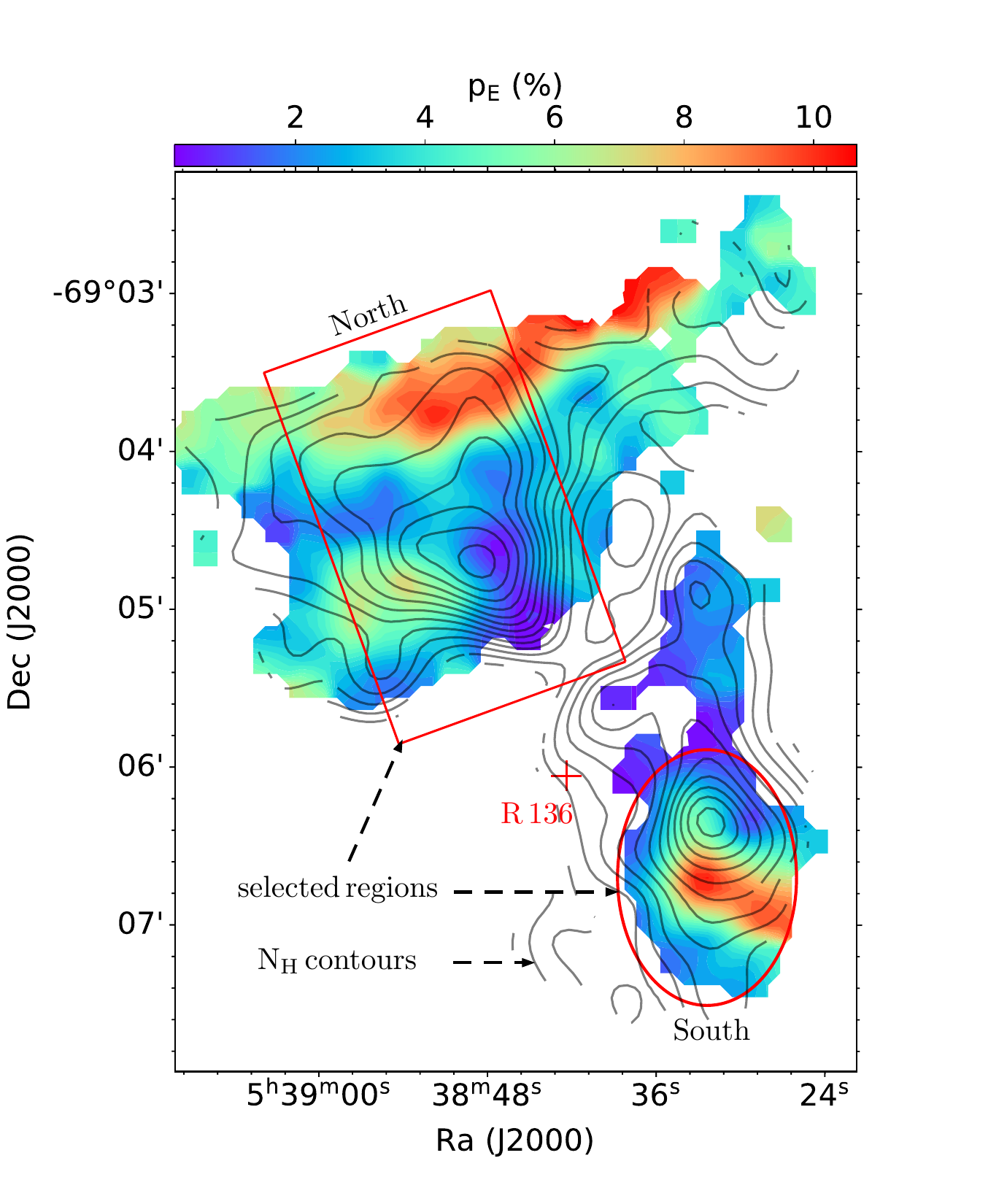}
    \caption{Sub-region selections. Background is the polarization degree at band E of the common space position which all SOFIA/HAWC+ detected data. The contours show the gas density map. The red rectangle (origin at 5$^{h}$38$^{m}$51.041$^{s}$ -69$^{\circ}$04$'$25.084$''$, width=17.12$^{s}$, height=2.5$'$) and ellipse (origin at 5$^{h}$38$^{m}$32.4$^{s}$ -69$^{\circ}$06$^{'}$42$^{''}$, major axis = 6.05$^{s}$, minor axis = 0.81$^{'}$) show the selected regions, where we are interested in.}
    \label{fig:selection_region}
\end{figure}
\begin{figure}
    \centering
    \includegraphics[width=0.45\textwidth]{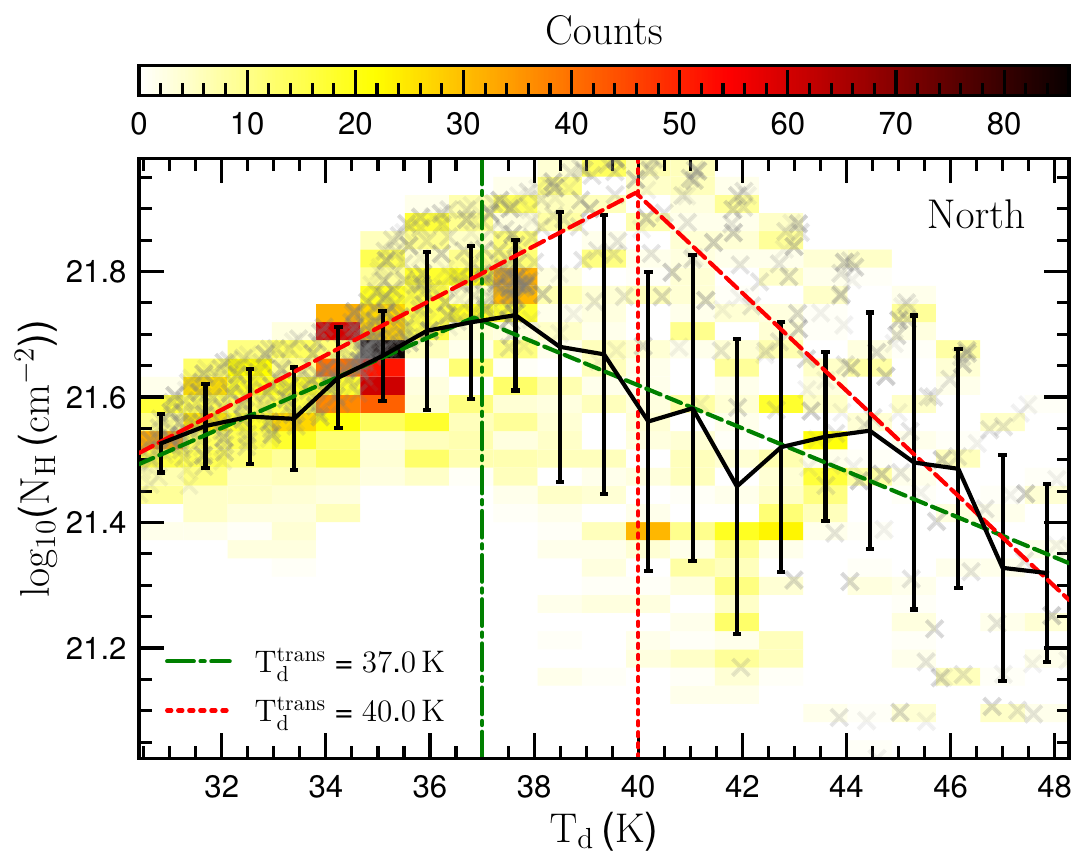}
    \includegraphics[width=0.45\textwidth]{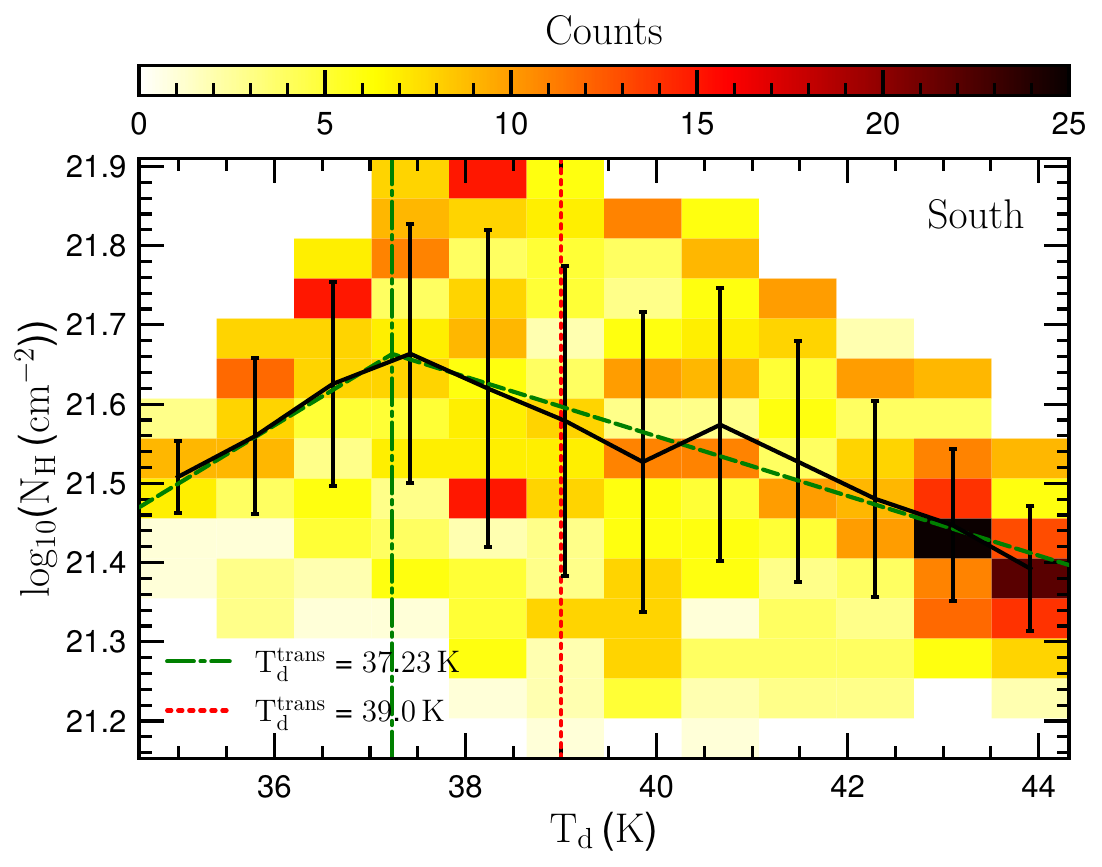}
    \caption{Relation of the gas column density and the dust temperature in the North region (upper panel) and in the South region (lower panel). They are positively correlated for $T_{\rm d}<37\K$, while they are negatively correlated for $T_{\rm d}>37\K$.}
    \label{fig:NH_Td}
\end{figure}
\subsection{Polarization degree versus dust temperature and gas column density}
For the optically thin case, the total dust emission intensity is $I_{\nu}\propto B(T_{\d})\tau_{\nu}\sim \kappa_{\nu}N_{\H}B(T_{\d}) \times R$ with $R$ the dust-to-gas ratio and $\kappa_{\nu}$ the dust opacity, which is the product of the gas column density and dust temperature along the line of sight. Therefore, the $p-I$ relation reflects the overall effect of both the radiation field and the gas density on grain alignment, assuming a uniform average magnetic field. To disentangle the effect of the radiation field from the gas density, we analyze the variation of the polarization degree with the dust temperature and the gas column density. The first analysis directly reveals the grain alignment and disruption by RATs, while the second one reflects the effect of grain randomization induced by gas collisions.

To make comparison among 3 bands, we first smooth the maps of the dust temperature, gas column density, and SOFIA/HAWC+ bands C and D to the $18.7''$ FWHM, which is equivalent to the SOFIA/HAWC+ band E beam-size. Then, we selected the common space position where three SOFIA/HAWC+ bands all detect data. The final map is shown in Figure \ref{fig:selection_region}. From this final map, we analyze the correlations in two sub-regions: within a rectangle in the North, and within an ellipse in the South. These extraction areas cover the representative features of both the gas density and the dust temperature in 30 Dor, which are shown in Figure \ref{fig:selection_region}.

\begin{figure*}
    \centering
    \includegraphics[width=1.0\textwidth]{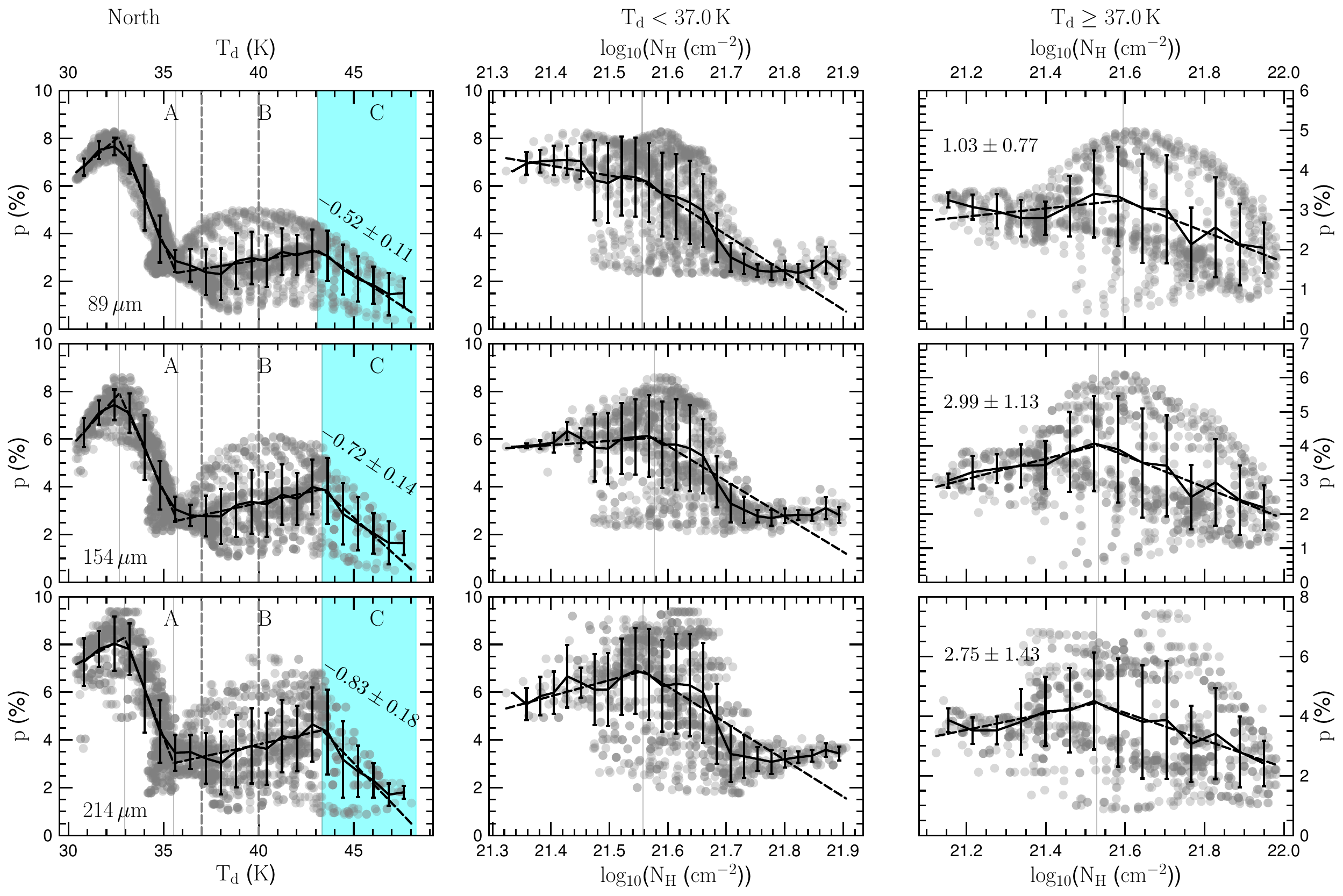}
     \caption{Left panel: correlation of the polarization degree to the dust temperature. Middle and right panels: correlation of the polarization to the gas column density within $T_{\rm}<37\K$ and $T_{\rm d}\geq 37\K$, respectively. Solid black line is the weighted-mean within $1-\sigma$ uncertainty in each bin. Dashed black line is the piecewise line fitting. On the left column, the dashed vertical lines indicate the corresponding $T^{\rm trans}_{\d}$. The $p-T_{\d}$ relation shows a three features: $p$ decreases (region A), slightly increases (region B), and decreases (region C) as $T_{\d}$ increases. The mean value of the slopes and their associated 95$\%$ level of confidences are given in region C.}
    \label{fig:p_vsTdNH_allnorth}
\end{figure*}
\begin{figure*}
    \centering
    \includegraphics[width=1.0\textwidth]{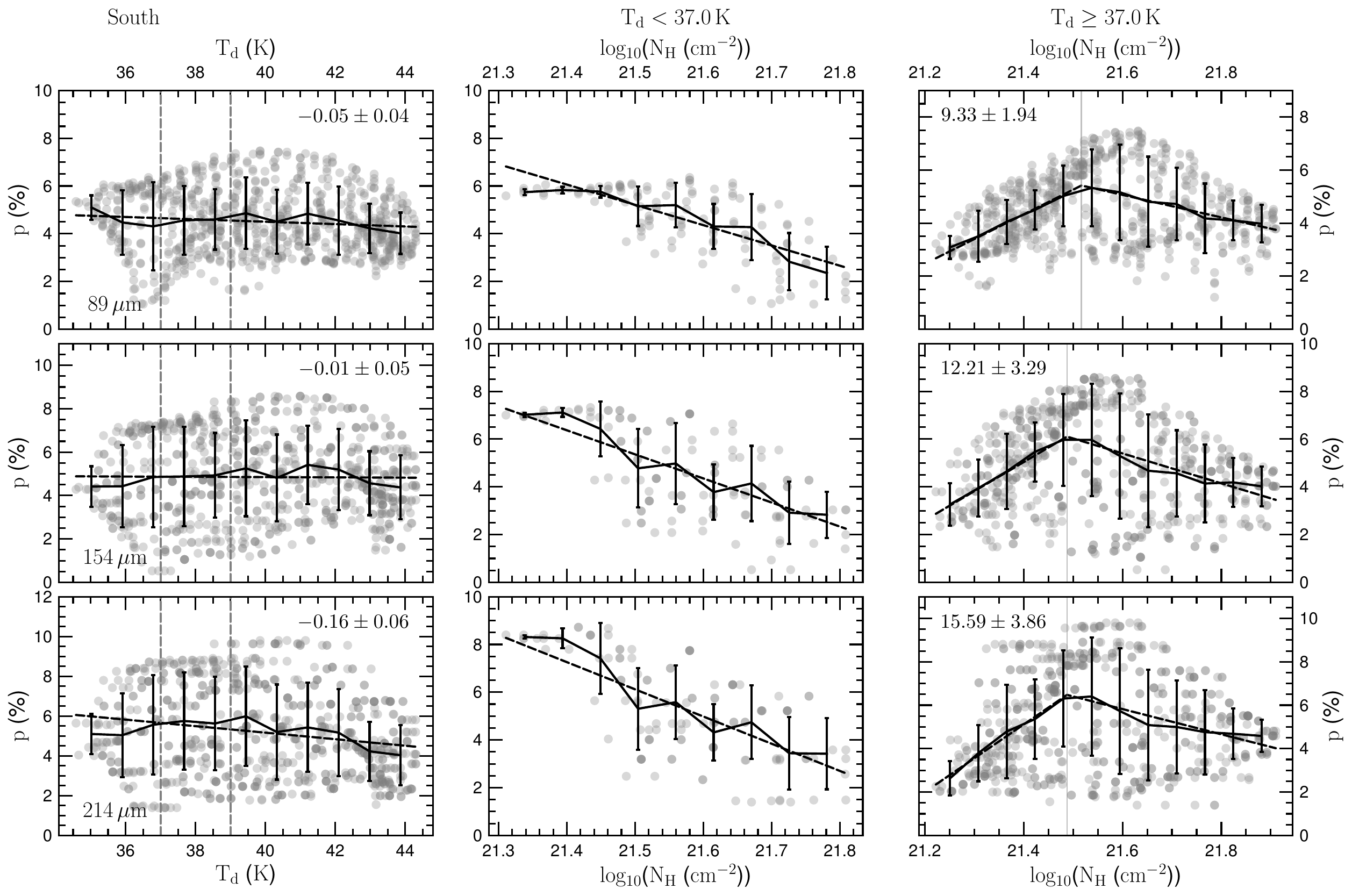}
    \caption{Similar to Figure \ref{fig:p_vsTdNH_allnorth} but for the South region. The polarization degree tends to decrease slowly with $T_{\d}$ (left panel). The polarization degree decreases with $N_{\H}$ for $T_{\d}<37\K$ (middle panel), but it increases and decreases with $N_{\H}$ for $T_{\d}>37\K$ (right panels).}
    \label{fig:p_vsTdNH_allsouth}
\end{figure*}

Figure \ref{fig:NH_Td} shows the two-dimensional histogram of the gas column density and the dust temperature relation for these two representative North (upper panel) and South (panel) regions. In the North, a \textsc{pwlf} fitting (\citealt{pwlf}) to all data points shows the gas column density is positively proportional and then negatively proportional to the dust temperature, for which the dust is below and above $T^{\rm trans}_{\rm d}\simeq 37\K$. However, since the error bar is significantly large, we fit an upper envelope of the distribution and obtain a slightly higher value of $T^{\rm trans}_{\rm d}\simeq 40\K$. In the South, the transition temperature occurs around $37-39\K$. However, the plot is much more scattered then in the North.

In Figure \ref{fig:p_vsTdNH_allnorth}, we show the relations of the polarization degree in 3 bands to the dust temperature (left panels from top to bottom). The black line shows the weighted mean fit (filled dots), and the black dashed line shows the piecewise line fit to the data. The separation owing to the $T_{\d}-N_{\H}$ is shown by two vertical dashed lines. The corresponding relations to the gas column density are shown in the middle and right panels, respectively. Notably, we only show the separation for $T^{\rm trans}_{\rm d}\simeq 37\K$ here because the behavior is the same for any values in between $37\K$ and $40\K$. 

For $T_{\d}\leq 37\K-40\K$, in which the dust temperature and the gas column density both increase toward the R$\,$136, the polarization degree rapidly increases, then decreases and finally gets nearly plateau with increasing dust temperature (region A in left panels). Meanwhile, the polarization degree tends to decrease up to $N_{\H}\sim 10^{21.7}\,\cm^{-2}$ and gets plateau with increasing the gas column density (middle panels). 

For $T_{\d}> 37\K-40\K$, in which the dust temperature and the gas column density are anti-correlated toward the R$\,$136, the polarization degree first increases slightly and then decreases with increasing both the dust temperature (region B and C in left panels) and the gas column density (right panels). Furthermore, the declining slope of $p-T_{\d}$ is steeper for the longer wavelength, i.e., the mean slope of $-0.52$ at $89\,\mu$m, $-0.72$ at $154\,\mu$m and $-0.83$ at $214\,\mu$m.

Similarly, Figure \ref{fig:p_vsTdNH_allsouth} shows the correlations of the polarization degree to the dust temperature and the gas column density in the South region. Although these plots are scattered, one can see a weak correlation between the polarization degree with the dust temperature.
The relation to the gas column density is also separated for the different range of dust temperature according to Figure \ref{fig:NH_Td}. The relation to the gas column density looks somehow similar to the middle and right panels of Figure \ref{fig:p_vsTdNH_allnorth}. For $T_{\rm d}<37-39\K$ (middle panel), the polarization degree shows a monotonic decrease with increasing the column density. However, for $T_{\rm d}\geq 37-39\K$ (right panel), the polarization degree first increases and then decreases with increasing the column density, as seen in the North region.

In summary, the $p-T_{\d}$ and $p-N_{\H}$ relations are not monotonic as in $p-I$. The most interesting feature is the anti-correlation of $p$ vs. $T_{\d}$ and the correlation of $p$ with $N_{\H}$, which cannot be explained by the basic RAT alignment theory alone.

\section{Discussion}\label{sec:discussion}
In this section, we will discuss the implications of the observed $p-I$, $p-T_{\d}$, and $p-N_{\H}$ relations for the leading theory of grain alignment and disruption based on RATs.

\subsection{On the $p-I$ relation and grain alignment} \label{sec:grain_align}
The variation of the polarization degree of thermal dust emission with the total emission intensity ($I$) is a popular analysis of polarimetric data, which provides information on grain alignment and magnetic fields. Numerous studies show that the polarization decreases with $I$ as $p\propto I^{-\alpha}$ with the slope $\alpha \simeq 0 - 1$. If grain alignment and the magnetic field are uniform throughout the cloud, one expects $\alpha=0$. If grain alignment only occurs in the outer layer and becomes completely lost in the inner region, one expects $\alpha=1$ (\citealt{2008ApJ...674..304W}). The latter slope is previously reported in the case of starless cores (see e.g., \citealt{2004ApJ...600..279C}; \citealt{2014A&A...569L...1A}; \citealt{2015AJ....149...31J}). Theoretically, for uniform grain alignment in the cloud, the stochastic magnetic field is found to induce a slope of $\alpha= 0.5$ (see \citealt{1992ApJ...389..602J,2015AJ....149...31J}).

\begin{figure*}    
\centering
    \includegraphics[width=0.9\textwidth]{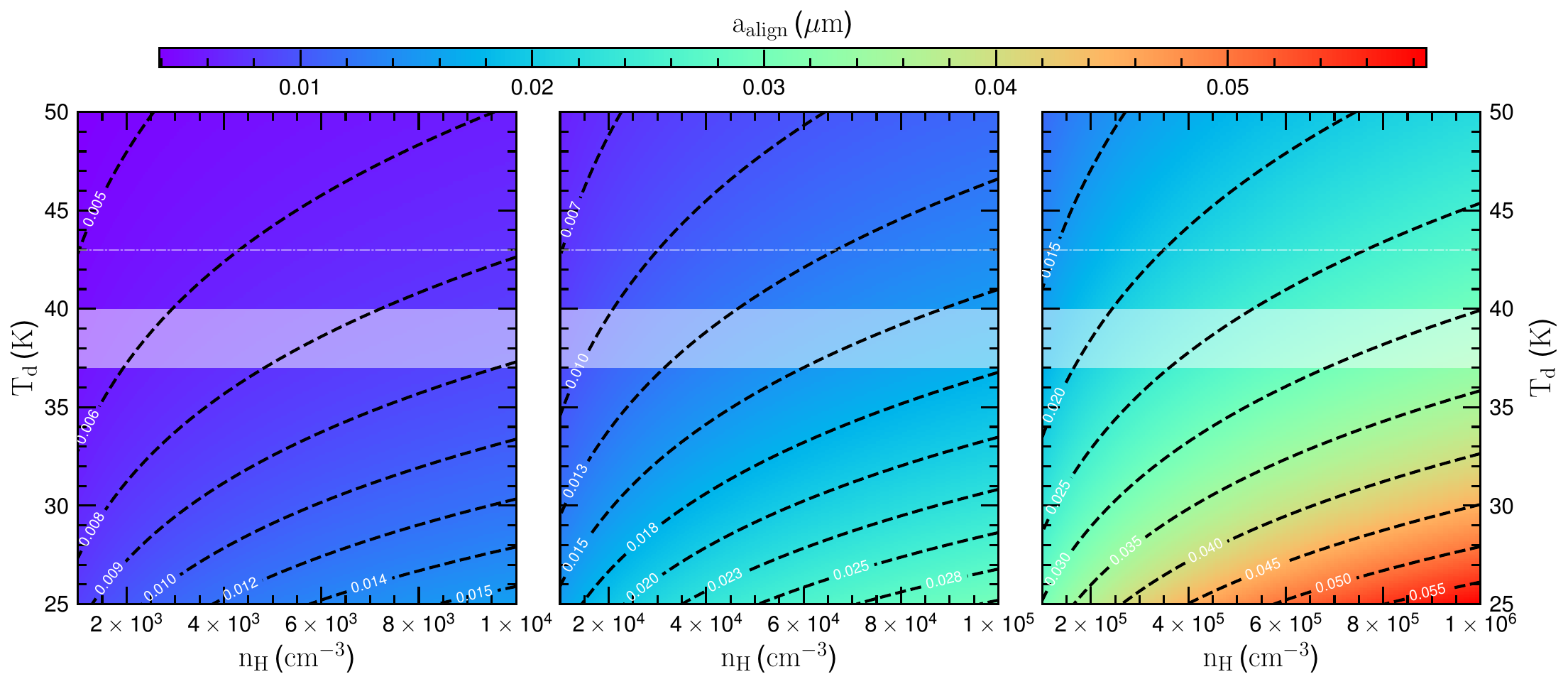}
    \caption{The alignment size (minimum size of aligned grains by RATs, $a_{\rm align}$) as functions of the dust temperature ($T_{\d}$) and the gas volume density ($n_{\H}$) computed with $\bar{\lambda}=0.3$, $\gamma=1$, and $T_{\rm gas}=T_{\d}$. The alignment size is larger (smaller) for higher (lower) gas density and lower (higher) dust temperature. The white region shows the range of the transition temperature ($T^{\rm trans}_{\d}$).}
    \label{fig:align}

    \includegraphics[width=0.75\textwidth]{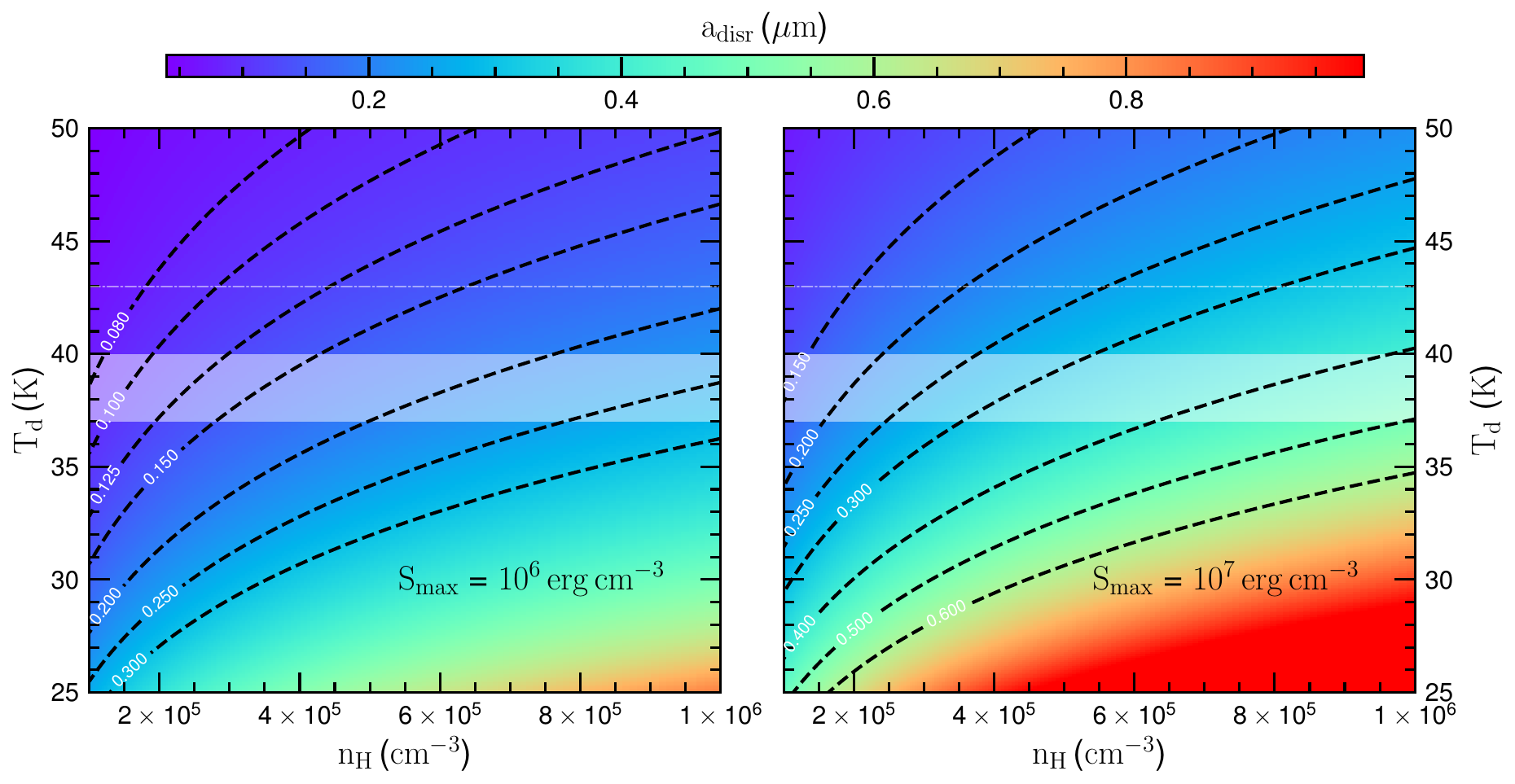}
    \includegraphics[width=0.75\textwidth]{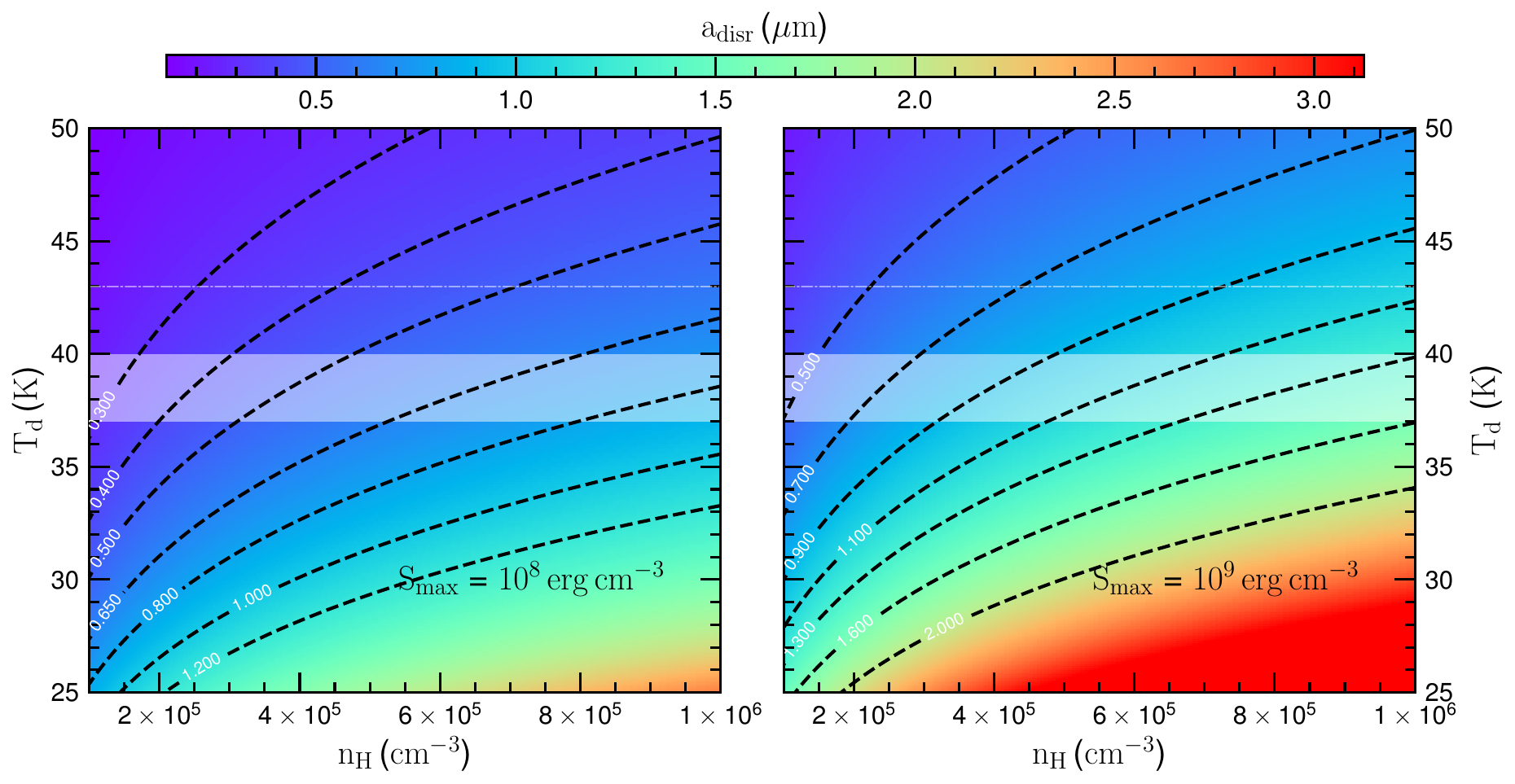}
    \caption{Similar to Figure \ref{fig:align} but for the grain disruption size ($a_{\rm align}$) for different values of the tensile strength ($S_{\rm max}$). $a_{\rm disr}$ is smaller (larger) for higher (lower) $T_{\d}$ and lower (higher) $n_{\H}$. For the same physical condition, $a_{\rm disr}$ is larger for higher $S_{\rm max}$. In each panel, the white dashed horizontal line represents to $T_{\d}=43\K$.}
    \label{fig:adisr}
\end{figure*}

In Section \ref{sec:p_vsI}, we showed the $p-I$ diagrams in all 3 bands in both North and South regions. In the North, a single power-law fitting shows that the diagram at 89$\,\mu$m has a steep slope of $\alpha>0.5$, while the ones at longer wavelengths show a shallower slope of $\alpha<0.5$. The slope of $p-I$ relations seem rather low compared to other regions, e.g., Auriga-California cloud ($\alpha \simeq 0.82$, \citealt{2021ApJ...908...10N}), Perseus B1 ($\alpha \simeq 0.8$, \citealt{2019ApJ...877...88C}), Ophiuchus B, C ($\alpha \simeq 0.6-0.7$, \citealt{2019ApJ...880...27P}), Serpens South ($\alpha \simeq 0.55$, \citealt{2020NatAs...4.1195P}). A double power-law fitting indicates the diagrams change from the first, steep slope of $\alpha>0.5$ to the second, shallower slope of $\alpha<0.5$ toward the peak dust emission intensity. This transition feature seems similar to the case of NGC 6334 (\citealt{2020arXiv201213060A}). In both cases, the shallow slope of the $p-I$ diagram is shown, which can be explained by the high grain alignment efficiency by RATs owing to the intense radiation flux from the luminous source R$\,$136. 

Indeed, according to the RAT alignment theory, the degree of dust polarization is determined by the minimum size of aligned grains ($a_{\rm align}$). A smaller value of $a_{\rm align}$ results in a higher $p$ because the size distribution of aligned grains (i.e., from $a_{\rm align}$ to the maximum grain size ($a_{\max}$)) is broader. Similarly, a larger value of $a_{\rm align}$ causes the lower polarization due to a narrower size distribution of aligned grains, assuming that $a_{\max}$ is fixed (without RATD). The alignment size depends on the local physical conditions as $a_{\rm align} \sim n^{2/7}_{\H}T^{-12/7}_{\d}$ (see Equation \ref{eq:aalign}). 

Since 30 Dor is a complex region, it is very difficult to accurately constrain the local physical properties such as $n_{\H}$ and $T_{\d}$ (see \citealt{2019A&A...621A..62O}). 
Therefore, in Figure \ref{fig:align}, we show the map of $a_{\rm align}$ with respect to $T_{\rm d}$ and $n_{\H}$. The values of $n_{\H}$ span broadly from $10^{3}$ to $10^{6}\,\cm^{-3}$ (see \citealt{2016A&A...590A..36C} and \citealt{2019A&A...628A.113L}), while that of $T_{\d}$ is adopted from observations (Figure \ref{fig:I_TdNH}d). The alignment size increases with increasing gas density and decreasing dust temperature (from upper left corner to lower right corner). For a given dust temperature, $a_{\rm align}$ becomes larger for higher gas density (from left panel to right panel), which arises from the increase of the collisional disalignment rate with the gas density.

In the case of two slopes, when approaching the luminous source of R$\,$136, the local dust temperature (i.e., radiation strength) increases, which significantly decreases $a_{\rm align}$ (see Figure \ref{fig:align}). As a result the polarization degree of thermal dust emission increases (\citealt{2021ApJ...908..218H}). This can reproduce the shallow slope of $\alpha\sim 0.3-0.4$ observed toward the peak intensity. For the first steep slope of $\alpha>0.5$, the alignment size increases due to the decrease of the interstellar radiation field as well as (the collisional damping becomes more significant), and one expects the slope of $\alpha=1$ when grain alignment is completely lost. However, if grain growth occurs, the size distribution of aligned grains ($a\sim a_{\rm align}-a_{\rm max}$) is still finite, which can produce the slope shallower than $\alpha\sim 1$ (see \citealt{2021ApJ...908..218H}). Incidentally, the optical to near-infrared observations of 30 Dor by {\it Hubble Space Telescope} (\citealt{2014MNRAS.445...93D}) reported a high value of the ratio of the total and selective extinction $R_{\rm V}$ as 4.4, which implies the grain growth in 30 Dor. Other effects, including the tangling of the magnetic fields, can also reproduce the first slope.

In the South, the $p-I$ diagrams illustrate two clusters with different properties. The first group shows a quite shallow slope with $\alpha \simeq 0.2-0.5$ at around the peak of the intensity and gas column density, which is opposite to the North. The second group otherwise shows a very deep slope as $\alpha \simeq 1$. Naturally, the contradiction of the polarization degree reflects a difference in the radiation flux and the grain size distribution. However, the dust temperature is not the highest at the dense gas (Figure \ref{fig:I_TdNH}d). Thus, it is likely that grains are larger in the dense region and responsible for a shallow slope of the $p-I$ diagram. 

The North and South regions share a common feature that the polarization degree seems to be larger at greater distances from the central source with a low gas density and low dust temperature. An interesting feature is that the polarization degree appears to be low in the vicinity of R$\,136$, in which the radiation is strong and the gas density is not the densest (Figure \ref{fig:I_TdNH}c). This is evidence in support of the RATD mechanism, which will be discussed in detail in the next section.

\begin{figure*}
    \centering
    \includegraphics[width=0.9\textwidth]{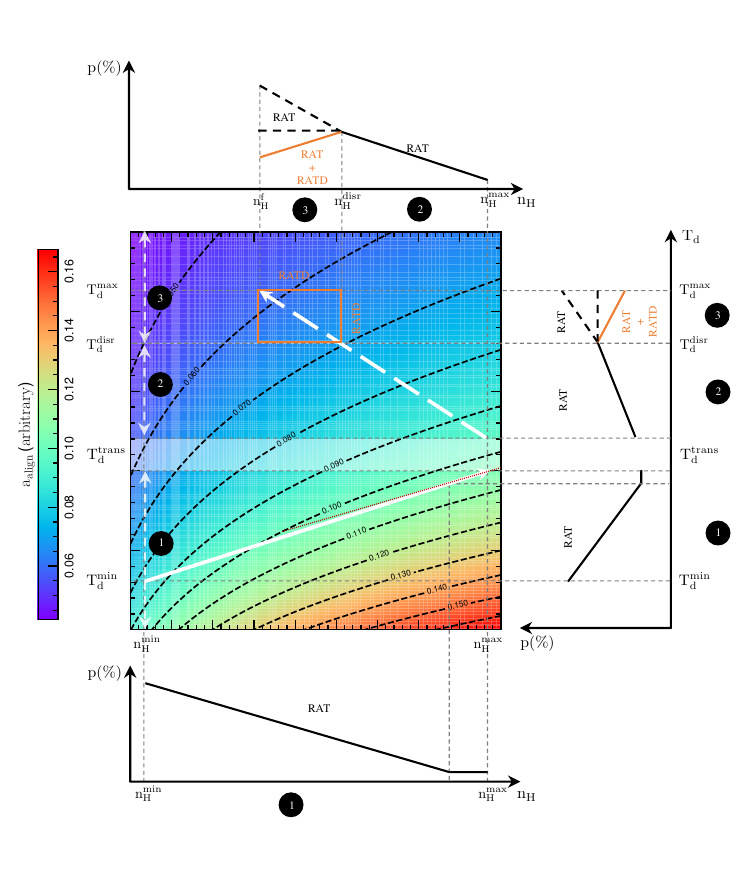}
    \caption{Understanding the thermal dust polarization in 30 Dor in the light of RAT alignment and RATD. According to the RAT alignment theory, the polarization degree is determined by the grain alignment size ($a_{\rm align}$), which depends on the dust temperature ($T_{\d}$) and the gas volume density ($n_{\H}$). Therefore, the observed $p-T_{\d}$ and $p-N_{\H}$ trends could be explained by three distinct stages denoted by $\textcircled{1}$, $\textcircled{2}$, and $\textcircled{3}$. For \textcircled{1} with $T_{\d}<T^{\rm trans}_{\d}$, $T_{\d}$ is positively proportional to $n_{\H}$ (solid white arrow), the grain alignment size increases along this direction only if $n_{\H}$ increases faster than the radiation strength $U\propto T_{\d}^{6}$, which results in the decrease of the polarization degree ($p$) with both $T_{\d}$ and $n_{\H}$. If $T_{\d}$ increases such that $U$ increases faster than $n_{\H}$, one expects $p$ to go up. For $\textcircled{2}$ with $T_{\d}>T^{\rm trans}_{\d}$, $T_{\d}$ is negatively proportional to $n_{\H}$ (dashed white arrow), $p$ increases with increasing $T_{\d}$ and decreasing $n_{\H}$ as $a_{\rm align}$ becomes smaller along this direction. Finally, for $\textcircled{3}$ with $T_{\rm d}>T^{\rm disr}_{\d}$, the RATD effect becomes active, resulting in the drop of $p$ toward higher $T_{\d}$ and lower $n_{\H}$ (orange lines), which explains the observed trend. Without RATD, $p$ is expected to continue increasing or be constant. Therefore, the  joint effect of grain alignment and disruption by RATs can successfully explain the observed polarized thermal dust emission of 30 Dor.}
    \label{fig:pol_sketch}
\end{figure*}
\subsection{On the $p-T_{\d}$ and $p-N_{\H}$ relations and grain alignment}
As shown in Figure \ref{fig:p_vsTdNH_allnorth}, the variation of $p$ with $T_{\d}$ and $N_{\H}$ in the North region is more complex than the $p-I$ relation. For $T_{\d}\leq 37\K$, $p$ decreases with increasing $T_{\d}$ (left panels) and $N_{\H}$ (middle panels). For $T_{\d}>37\K$, $p$ first increases with increasing $T_{\d}$ and $N_{\H}$, then decreases when $T_{\d}$ becomes sufficiently large (see B and C regions in the left panels). The increase of $p$ with $T_{\d}$ and the decrease of $p$ with $N_{\H}$ are expected from the RAT alignment theory as the alignment size decreases with increasing $U$ and decreasing $n_{\H}$ (see Figure \ref{fig:align}, also Equation \ref{eq:aalign}). However, the decrease of $p$ with $T_{\d}$ is unexpected from the basic RAT alignment theory because the alignment size $a_{\rm align}$ is smaller for larger $T_{\d}$ ($U$). Similarly, the increase of $p$ with $N_{\H}$ is unexpected because $a_{\rm align}$ increases with the gas density.

In the South, the relations of $p$ to $T_{\rm d}$ and $N_{\H}$ are more scattered and not very obvious as shown in Figure \ref{fig:p_vsTdNH_allsouth}. However, the South is likely closer to R$\,$136 than the North, and the selected area covers mostly the region where the grain alignment efficiency is high. However, the polarization degree seems to not vary with $T_{\d}$ and decreases with decreasing the gas column density for $T_{\d}>T^{\rm trans}_{\d}$. Same as in the North, the latter trend is opposite to the expectation by the basic RAT alignment theory.

\subsection{A new evidence of the RATD mechanism}

In Figure \ref{fig:p_vsTdNH_allnorth} and \ref{fig:p_vsTdNH_allsouth}, we found that the polarization degree first increases and then decreases as the dust temperature increases and the gas column density decreases (region C in Figure \ref{fig:p_vsTdNH_allnorth}) for $T_{\rm d}>43\K$. The decrease of $p$ with $T_{\d}$ (higher radiation flux, closer R$\,$136) and the increase of $p$ with decreasing $N_{\H}$ cannot be explained by the basic RAT alignment theory, as discussed in the previous subsection, unless the increasing dust heating is associated with the increase in the tangling of the magnetic field. Below, we suggest that the combination of the RAT alignment with the RATD mechanism (\citealt{2019NatAs...3..766H}) can reproduce this feature.

According to the RATD mechanism, dust grains subject to a strong radiation field (i.e., close the source R$\,$136) can be spun-up to extremely fast rotation. The induced centrifugal stress due to grain suprathermal rotation can exceed the tensile strength of the grain material, resulting in the spontaneous fragmentation of large grains into smaller ones. The depletion of large grains causes the decrease of the thermal dust polarization degree at long wavelengths (see \citealt{2020ApJ...896...44L}; \citealt{2021ApJ...906..115T}).  

To study if RATD can reproduce the observed anti-correlation of $p-T_{\d}$, let us estimate the critical size above which dust grains are disrupted, $a_{\rm disr}$. The disruption size depends on the local conditions and the grain structure as $a_{\rm disr} \sim n_{\H}^{1/2}T^{-3}_{\rm d}S^{1/4}_{\rm max}$ (see Section \ref{sec:App_adisr}). As shown in \cite{2021ApJ...906..115T,2021ApJ...908..159T}, the disruption occurs at the critical disruption temperature of $T^{\rm disr}_{\d}\simeq 32\K$ for $n_{\H}\simeq 10^{5}\,\cm^{-3}$, and $T^{\rm disr}_{\d} \simeq 71\K$ for $n_{\H} \sim 10^{6}\,\cm^{-3}$. Thus, the gas volume density in the region within $43\leq T_{\d}<50\K$ could be in $(10^{5}-10^{6}\,\cm^{-3})$ interval. Figure \ref{fig:adisr} shows the map of the grain disruption size as a function of $T_{\d}$ and $n_{\H}$ for different value of $S_{\rm max}$. Grains within composite structure ($S_{\rm max}\sim 10^{5}-10^{7}\,\erg\cm^{-3}$) are being disrupted at smaller size than the more compact grains (higher value of $S_{\rm max}$). For instance, for $n_{\H}=6\times 10^{5}\,\cm^{-3}$ and $T_{\d}=43\K$, $a_{\rm disr}$=0.15, 0.26, 0.46, 0.82$\,\mu$m for $S_{\max}=10^{6},\,10^{7},\,10^{8},\,10^{9}\,\erg\cm^{-3}$, respectively. Thus, the RATD mechanism can reasonably disrupt the composite grains for $T_{\d}\geq 43\K$. The disruption size gets smaller for higher $T_{\d}$.

Figure \ref{fig:pol_sketch} sketches our understanding on the main feature of dust polarization in 30 Dor expected from theoretical modeling of dust polarization based on grain alignment and disruption by RATs (\citealt{2020ApJ...896...44L}). For an illustration, a color map of $a_{\rm align}$ (in arbitrary units) is shown as functions of $T_{\d}$ and $n_{\H}$. 
The feature of $p$ fundamentally depends on the $T_{\d}-n_{\H}$ relation. If we assume $n_{\H} \sim N_{\H}$ maps (e.g., the local variation of the depth of cloud is significantly small or negligible), the observed polarization degree ($p$) profile could be explained in three ranges of $T_{\d}$. 

For $T_{\d}<T^{\rm trans}_{\d}$ (at which the $T_{\d}-n_{\H}$ relation changes the sign from positive to negative, see Figure \ref{fig:NH_Td}), $T_{\d}$ positively correlates to $n_{\H}$. If the gas density increases faster than the radiation strength ($U\propto T_{d}^{6}$), such that the rotational damping by gas collisions is sufficiently strong to prevent the grains from being spun-up to the disruption limit by RATs, $a_{\rm align}$ is larger with increasing $T_{\d}$ and $n_{\H}$ (solid white arrow). This increment of $a_{\rm align}$ consequently causes the polarization degree to decrease to both the dust temperature and gas density as expected from the RAT alignment theory. In this case, $a_{\rm align}$ might also become constant with a specific range of $T_{\d}$ and $n_{\H}$ because of their opposite effect. These features explain the observed region A of $p-T_{\d}$ in Figure \ref{fig:p_vsTdNH_allnorth} and $p-N_{\H}$ in the middle panel of Figures \ref{fig:p_vsTdNH_allnorth}, \ref{fig:p_vsTdNH_allsouth}. In the case if the dust temperature increases such that $U$ increases faster than the gas density, the rotational damping by gas collisions is insufficiently strong to prevent the grains from being spun-up by RATs, $a_{\rm align}$ is getting smaller as $T_{\d}$ and $n_{\H}$ increase. Thus, the polarization degree is increasing instead of decreasing, which contradicts to the observed trend of 30 Dor.

For $T_{\d}>T^{\rm trans}_{\d}$, $T_{\d}$ and $n_{\H}$ are anti-correlated (see Figure \ref{fig:NH_Td}). The gas collision damping is further weak, the grain alignment size gets smaller as $T_{\d}$ increases (dashed white arrow). Thus, the polarization degree gets increased toward higher $T_{\d}$ and lower $n_{\H}$. This feature could explain the region B of the observed $p-T_{\d}$ in Figure \ref{fig:p_vsTdNH_allnorth}.

For $T_{\d}>T^{\rm disr}_{\d}$, the depletion of large grains due to the RATD mechanism results in the decrease of the polarization degree toward higher $T_{\d}$ and lower $n_{\H}$. This feature successfully reproduces the observed $p-T_{\d}$ in region C of Figure \ref{fig:p_vsTdNH_allnorth} and $p-N_{\H}$ in the right panels of Figures \ref{fig:p_vsTdNH_allnorth}, \ref{fig:p_vsTdNH_allsouth}. Otherwise, the polarization degree is expected to continuously increase or at least flat accordingly to the RAT alignment theory (see \citealt{2020ApJ...896...44L}).

However, there is a question remaining why the relation of $p$ to $T_{\d}$ is weak in the South. This slow variation may be due to the saturation of RATD, which occurs when $T_{\d} >> T_{\rm disr}$ (see Figure 14 in \citealt{2020ApJ...896...44L}). Induced small grains likely have compact structures and RATD ceases to act at some high $T_{\d}$. Furthermore, when $T_{\d}$ is large, IR damping is so strong that $a_{\rm disr}$ becomes to slowly change with $T_{\d}$. Both effects induce the slow variation of $p$ with $T_{\d}$.

Previous studies have reported the anti-correlation of the dust polarization with grain temperature for several molecular clouds (\citealt{2020A&A...641A..12P}), the $\rho$ Ophiuchus A which is irradiated by a nearby B-association star (\citealt{2019ApJ...882..113S}; \citealt{2021ApJ...906..115T})\footnote{In the case of $\rho$ Ophiuchus A, the observed trend of $p-T_{\d}$ is explained by the processes \textcircled{2} and \textcircled{3} in Figure \ref{fig:pol_sketch}.}, and at 89, 154, 214, and 850$\,\mu$m of a cloud surrounding the massive star BN/KL (\citealt{2021ApJ...908..159T}). Numerical modeling studies (\citealt{2020ApJ...896...44L}) show that the RATD mechanism (\citealt{2020Galax...8...52H}) could successfully reproduce the observed anti-correlation of the polarization with grain temperature at the same FIR and submm wavelength ranges. The observed anti-correlation of $p-T_{\d}$ in 30 Dor is a new evidence for the RATD mechanism.

\subsection{Dust polarization mechanisms} 
Our analysis so far is focused on the polarization degree with the assumption that far-IR dust polarization is induced by dust grains aligned with a preferred direction. Modern theory of RAT alignment theory implies that such a preferred axis could be the magnetic field (B-RAT) or the radiation direction (k-RAT) (\citealt{2007MNRAS.378..910L}). As a result, the polarization vectors of polarized thermal emission are perpendicular to the magnetic field (radiation direction). The B-RAT alignment is most likely in the diffuse ISM and molecular clouds, whereas the k-RAT can occur near a strong radiation source where the rate of grain precession around the radiation direction is larger than the Larmor precession around the magnetic field (\citealt{Hoang:2016}; \citealt{LazarianHoang:2019}). In this case, the polarization vectors of thermal dust emission are perpendicular to the radiation direction. Therefore, the polarization vectors provide valuable constraints on dust magnetic properties and alignment mechanisms (\citealt{Chuss:2019}; \citealt{LazarianHoang:2019}; \citealt{Pattle:2021}). 

In the South region of Figure \ref{fig:hawc_map}, the polarization vectors (the E-vectors) appear to be concentric with respect to the intense source R$\,$136. Such a polarization pattern is different from the one in the North region which is far from the source R$\,$136 where grains are most likely aligned via B-RAT. This indicates that the alignment mechanisms in the South may be due to the alignment with the radiation direction. Evidence for k-RAT is previously reported for some star-forming regions (\citealt{Chuss:2019}; \citealt{Pattle:2021}) and protoplanetary disks (\citealt{Kataoka2017}; \citealt{Stephens2017}). We will study in detail this issue in a follow up paper on 30 Dor.

\section{Summary}\label{sec:summary}
In this paper, we report our results analyzing the far-IR polarimetric observations of 30 Doradus by SOFIA/HAWC+. 30 Dor cloud is located in the Large Magellanic Cloud and mainly irradiated by an intense radiation source from the massive R$\,$136 cluster. 
Thus, this region is an ideal target to study the physics of grain alignment and disruption driven by RATs. 
To infer the alignment and disruption properties of dust grains, we have performed three different analyses based on the polarization degree, including $p-I$, $p-N_{\H}$, and $p-T_{\d}$. Our main findings are summarized as bellows

\begin{itemize}
    \item [1] The thermal dust polarization varies across the 30 Dor cloud, in which there are two main regions, namely North and South in relative to the massive star cluster R$\,$136. In the North region, the polarization patterns are likely radial from R$\,$136 within the polarization angle peaked at $\simeq 20^{\circ}$, while they are concentric around R$\,$136 within the peaked polarization angle of $\simeq -60^{\circ}$ in the South region.
    
    \item [2] 
    For the North region, a single power-law fitting show a shallow slope of the $p-I$ diagrams, which could be explained by an intense radiation field. A double power-law fitting shows a change from a steeper slope of $\alpha>0.5$ to a shallower slope as $\simeq I^{-0.4} - I^{-0.3}$ toward higher intensity. The latter shallow slope most likely arises from the enhanced alignment of grains by RATs when approaching the source R$\,$136. 
    
    \item[3]
The variations of the polarization degree with the dust temperature and gas column density in the North are complex. The polarization degree decreases with increasing the dust temperature and the gas column density for $T_{\d}<37\K$, which might result from the loss of grain alignment due to strong collisional damping by denser gas. For $T_{\d}>37\K$, the polarization degree increases as the dust temperature increases and the gas column density decreases. However, the polarization degree decreases as the dust temperature increases beyond $\sim 43\K$ and the gas column density further decreases. Such an anti-correlation of $p-T_{\d}$ is consistent with the prediction of the RATD theory that the dust polarization decreases due to disruption of large grains toward the intense radiation source R$\,$136.

    \item [4] 
        In the South region, the polarization degree is higher, and the $p-I$ slope is shallower at the peak of the intensity and gas column density. Whereas, the polarization degree is the highest at further away (low intensity and less dense gas) and decreases steeply as $p\propto I^{-1}$ backward to R$\,$136 (higher intensity and denser gas). The first feature is likely caused by enhancement of grain alignment by RATs due to the radiation from the source R 136. The second feature is unexpected from the basic RAT alignment theory, but is consistent with the prediction by the RATD effect, as in the North.
         
    \item [5] 
        In the South, the stronger grain randomization by gas collisions in the densest part results in a weak correlation of the polarization degree to the dust temperature. Whereas, the polarization degree likely decreases toward low and high gas column density. However, the relations are quite scattered and show a large error bar.
\end{itemize}

The values of the dust temperature and gas column density using in this work are derived from the modified black-body fitting to the \textit{Herschel} data, and they thus encompass the projection effect. Our conclusions are given without taking the B-field tangling into account. The variation of the magnetic field could result in the modification of the grain alignment and then the polarization degree. Despite these aspects, our SOFIA/HAWC+ data show that the grain alignment is complicate and significantly varied across the 30 Dor cloud, and evidence a strong impact of the intense radiation field of a massive star cluster R$\,$136. The drop of the polarization degree as close to the central source, in which the gas column density is not peaked, is contrary to the RAT alignment theory. The lower abundance of large grains via the disruption mechanism by radiative torque could be responsible for this declination.    

\textit{Acknowledgments}: We thank the anonymous referee for a helpful report. This research is based on observations made with the NASA/DLR Stratospheric Observatory for Infrared Astronomy (SOFIA). SOFIA is jointly operated by the Universities Space Research Association, Inc. (USRA), under NASA contract NNA17BF53C, and the Deutsches SOFIA Institut (DSI) under DLR contract 50 OK 0901 to the University of Stuttgart. Financial support for this work was provided by NASA through award 4$\_$0152 issued by USRA. T.H is funded by the National Research Foundation of Korea (NRF) grant funded by the Korea government (MSIT) through the Mid-career Research Program (2019R1A2C1087045).

\textit{Facility}: SOFIA

\appendix
\section{Grain alignment size}\label{sec:App_align}
Grains are effectively aligned only when they can  rotate suprathermally, i.e., the grain angular velocity ($\omega$) is greater than the thermal velocity ($\omega_{\rm T}$) (\citealt{2008MNRAS.388..117H}; \citealt{Hoang:2016}). The minimum alignment size ($a_{\rm align}$) above which grains are aligned can be computed as $\omega(a_{\rm align}) \equiv 3\times \omega_{\rm T}$. As exposed to a radiation source, the minimum $a_{\rm align}$ is determined analytically as (see Equation 26 in \citealt{2021ApJ...908..218H}) 
\bea \label{eq:aalign}
    a_{\rm align} &\simeq& 0.055 \hat{\rho}^{-1/7} \left(\frac{\gamma}{0.1}\right)^{-2/7}\left(\frac{n_{\H}}{10^{3}\rm cm^{-3}}\right)^{2/7} U^{-2/7} \\ \nonumber
    &&\left(\frac{T_{\rm gas}}{10\K}\right)^{2/7}\left(\frac{\bar{\lambda}}{1.2\rm \mu m}\right)^{4/7}(1+F_{\rm IR})^{2/7}~~~{\rm \mu m},
\ena
where $\gamma$ is the anisotropy degree, $\bar{\lambda}$ is the mean wavelength, and $U$\footnote{$U=u_{\rm rad}/u_{\rm ISRF}$, with $u_{\rm rad}=\int u_{\lambda}d\lambda$ the radiation energy density of the local radiation field and $u_{\rm ISRF}\approx 8.64\times10^{-13}\erg\,\cm^{-3}$ the radiation energy density of the local interstellar radiation field (\citealt{Mathis1983}).} is the strength of the radiation field, respectively. $n_{\H}$ and $T_{\rm gas}$ are the number density and temperature of the gas. $F_{\rm IR}$ is the ratio of the IR damping to the collisional damping rate. If we use $U\simeq (T_{\rm d}/16.4\K)^{6}$ (\citealt{2011piim.book.....D}), one can see that the alignment size depends on the local properties as $a_{\rm align} \sim n^{2/7}_{\H} T^{-12/7}_{\rm d}$. The alignment size decreases (increases) with decreasing (increasing) gas density and increasing (decreasing) dust temperature as shown in Figure \ref{fig:align}.

\section{Grain disruption size}\label{sec:App_adisr}
Large grains exposed to an intense radiation field can be spun-up to extremely fast due to RATs. The fast rotation induces a strong centrifugal stress which opposes the binding energy that holds up the grain structure (i.e., tensile strength). Once the centrifugal stress exceeds the tensile strength, the grain is spontaneously fragmented. This fragmentation is called Radiative torque disruption (RATD). The critical grain size above which grains will be disrupted by the RATD mechanism is
\bea \label{eq:adisr}
    a_{\rm disr} &\simeq& 0.22\hat{\rho}^{-1/4}\gamma^{-1/2}\left(\frac{n_{\H}}{10^{3}\cm^{-3}}\right)^{1/2}U^{-1/2} \left(\frac{T_{\rm gas}}{10\K}\right)^{1/4} \\ \nonumber
    && \times \left(\frac{\bar{\lambda}}{0.5\rm \mu m}\right)\left(\frac{S_{\rm max}}{10^{7}\erg \cm^{-3}}\right)^{1/4} (1+F_{\rm IR})^{1/2}\mum,
\ena
where $S_{\rm max}$ is the tensile strength which is determined by the internal structure of dust grains (see Equation 30 in \citealt{2021ApJ...908..218H}). The compact grains have a higher tensile strength and are stronger than the composite ones (see \citealt{2019ApJ...876...13H}). One can see that $a_{\rm disr} \sim n_{\H}^{1/2}T^{-3}_{\rm d}S^{1/4}_{\rm max}$, which illustrates that the disruption size also decreases (increases) with decreasing (increasing) gas density and increasing (decreasing) dust temperature. Moreover, at the same local condition, grains within a weaker structure are easier to be disrupted than the stronger one as shown in Figure \ref{fig:adisr}.

\end{document}